\documentclass{aa}

\usepackage{graphicx}
\usepackage{txfonts}
\usepackage[breaklinks, colorlinks, citecolor=blue]{hyperref}
\usepackage{natbib}
\begin{document} 

   \title{Faraday tomography of LoTSS-DR2 data: I. Faraday moments in the high-latitude outer Galaxy and revealing Loop III in polarisation\thanks{{The movie associated to Appendix B is available at \url{https://www.aanda.org} and also at \url{https://data.fulir.irb.hr/islandora/object/irb:106}}}}

   \titlerunning{Draft}
   \authorrunning{Erceg et al.}
  
   \author{Ana Erceg\inst{1,2},
          Vibor Jelić\inst{1},
          Marijke Haverkorn\inst{2},
          Andrea Bracco\inst{1,3},
          Timothy W. Shimwell\inst{4,5},
          Cyril Tasse\inst{6,7},
          John M. Dickey\inst{8},
          Lana Ceraj\inst{1},
          Alexander Drabent\inst{9},
          Martin J. Hardcastle\inst{10},
          Luka Turić\inst{1}
          }

   \institute{Ruđer Bošković Institute, Bijenička cesta 54, 10 000 Zagreb, Croatia\\
              \email{aerceg@irb.hr}
         \and
             Department of Astrophysics/IMAPP, Radboud University, P.O. Box 9010, 6500 GL Nijmegen, The Netherlands
         \and 
             Laboratoire AIM, CEA / CNRS / Universit{\'e} Paris-Saclay, 91191 Gif-sur-Yvette, France
         \and
             ASTRON, Netherlands Institute for Radio Astronomy, Oude Hoogeveensedijk 4, 7991 PD, Dwingeloo, The Netherlands
         \and
             Leiden Observatory, Leiden University, P.O. Box 9513, 2300 RA Leiden, The Netherlands
        \and
             GEPI \& USN, Observatoire de Paris, CNRS, Universi{\'e} Paris Diderot, 5 place Jules Janssen, 92190 Meudon, France 
        \and
            Centre for Radio Astronomy Techniques and Technologies, Department of Physics and Electronics, Rhodes University, Grahamstown 6140, South Africa
        \and
            School of Natural Sciences, Private Bag 37, University of Tasmania, Hobart, TAS, 7001, Australia
         \and
            Th{\"u}ringer Landessternwarte, Sternwarte 5, D-07778 Tautenburg, Germany
         \and
            Centre for Astrophysics Research, Department of Physics, Astronomy and Mathematics, University of Hertfordshire, College Lane, Hatfield AL10 9AB, UK
            }
   \date{Received 17 September 2021 / Accepted 22 February 2022}

 
  \abstract
   {Observations of synchrotron emission at low radio frequencies reveal a labyrinth of polarised Galactic structures. However, the explanation for the wealth of structures remains uncertain due to the complex interactions between the interstellar medium and the magnetic field. A multi-tracer approach to the analysis of large sky areas is needed.
   }
   {This paper aims to use polarimetric images from the LOFAR Two metre Sky Survey (LoTSS) to produce the biggest mosaic of polarised emission in the northern sky at low radio frequencies {(150 MHz)} to date. The large area this mosaic covers allows for detailed morphological and statistical studies of polarised structures in the high-latitude outer Galaxy, including the well-known Loop III region.}
   {We produced a 3100 square degree Faraday tomographic cube using a rotation measure synthesis tool. We calculated the statistical moments of Faraday spectra and compared them with data sets at higher frequencies (1.4 GHz) and with a map of a rotation measure derived from extragalactic sources. }
   {The mosaic is dominated by polarised emission connected to Loop III.  Additionally, the mosaic reveals an abundance of other morphological structures, mainly {narrow and extended} depolarisation canals, which are found to be ubiquitous.}
   {We find a correlation between the map of an extragalactic rotation measure and the LoTSS first Faraday moment image. The ratio of the two deviates from a simple model of a Burn slab \citep{burn66} along the line of sight, which highlights the high level of complexity in the magnetoionic medium that can be studied at these frequencies.
   }


   \maketitle
%

  \section{Introduction}\label{sec:intro}
In the last decade, radio astronomy at low frequencies has flourished as a result of major technical advancements and the development of a new generation of radio interferometers. One such instrument is the LOw-Frequency
ARray (LOFAR; \citealt{vanhaarlem13}), which aims to, among other scientific goals, provide a deep imaging survey of the entire northern sky. 
Its ongoing 120~--~168 MHz survey, the LOFAR Two-meter Sky Survey (LoTSS, ~\citealt{shimwell17, shimwell19}), focusses on exploring the formation and evolution of supermassive black holes, galaxies, clusters, and large-scale structures in the Universe and our Galaxy. The observations of polarised Galactic synchrotron emission obtained in this survey are of extremely high value for use in studies of the Galactic magnetic field and interstellar medium (ISM).

Polarised synchrotron emission associated with structures in the Milky Way is affected by Faraday rotation as it propagates through a web of a magnetised and ionised ISM. As it propagates, its linear polarisation plane is rotated by a wavelength-dependent angle ($\Delta\theta$) described by Faraday depth ($\Phi$) as follows:
\begin{equation}\label{eq:FD}
     \frac{\Delta\theta}{\mathrm{[rad]}}=\frac{\lambda^2}{\mathrm{[m^2]}}\frac{\Phi}{\mathrm{[rad \ m^{-2}]}} = \frac{\lambda^2}{\mathrm{[m^2]}}0.81 \int_{0}^{d} \frac{n_e}{\mathrm{[cm^{-3}]}} \frac{B_\parallel}{\mathrm{[\mu G]}} \frac{\mathrm{dl}}{\mathrm{[pc]}},
\end{equation}
where {$\lambda$ is the wavelength}, $n_e$ is the electron density, {and} $B_\parallel$ is the magnetic field component parallel to the line of sight (LOS). The integral is evaluated over distance ${\rm d}l$ from the source (at distance 0) to the observer (at distance $d$). The Faraday depth is positive or negative if the {average} parallel component of the magnetic field is pointing towards or away from the observer, respectively. 

In the simplest case of a background polarised source and a foreground Faraday rotating medium, the observed wavelength-dependent change of the polarisation angle is linear with $\lambda^2$. The Faraday depth of the observed polarised emission is then equivalent to the rotation measure (RM), defined by the slope of the observed linear relation between $\theta$ and $\lambda^2$ \citep[e.g.][]{manchester72, ferriere21}.

The observed polarised emission can be disentangled into components according to the amount of Faraday rotation experienced by using the {RM synthesis} technique \citep{burn66, brentjens05} For a given location in the sky, this technique gives us a Faraday spectrum, that is to say a distribution of the observed polarised emission in Faraday depth. If RM synthesis is applied over a sky area, we can study the morphology of the observed polarised emission at different Faraday depths. Using this technique, we performed the so-called Faraday tomography, which allowed us to analyse complex distributions of magnetised and ionised gas along a LOS.

Rotation-measure synthesis has limitations which are defined by the following three parameters \citep[][see eq. 61-63]{brentjens05}: {the maximum observable Faraday depth, $\Phi_\mathrm{max}$, which is inversely proportional to the channel width of the observations, $\delta \lambda^2$; the resolution in Faraday depth, $\delta\Phi$, which is inversely proportional to the wavelength range the observation covers, $\Delta \lambda^2$; the largest scale in Faraday depth to which the observations are sensitive, $\Delta\Phi_\mathrm{max}$, which is inversely proportional to the minimum wavelength of the observation, $\lambda^2_\mathrm{min}$}. The wavelength dependence of both the resolution and the Faraday rotation (proportional to $\lambda^2$, see Eq.\ref{eq:FD}) makes the power of RM synthesis at low radio frequencies twofold. We can study observed emission at high resolution in Faraday depth (up to 1 $\mathrm{rad \ m^{-2}}$) and detect low column densities of magnetised and ionised gas. This is not possible at high radio frequencies ($\gtrsim 1~{\rm GHz}$) since the resolution is a few orders of magnitude lower than at low radio frequencies  \citep{jelic15, vaneck18}.

The structures seen {in the Faraday spectrum} can be either Faraday thin or thick {\citep{brentjens05}}. {A structure is considered Faraday thin if $\lambda^2\Delta\Phi \ll 1$, where $\Delta\Phi$ is the extent of the source in Faraday depth. On the other hand, if $\lambda^2\Delta\Phi \gg 1$, the structure is called Faraday thick.} Faraday thickness is wavelength-dependent, meaning that a structure, which is Faraday thick at low frequencies, becomes Faraday thin at higher frequencies.

Thus far, a number of authors have used RM synthesis to analyse Galactic polarised synchrotron emission in seven single field observations with LOFAR \citep{iacobelli13, jelic14, jelic15, vaneck17, turic21}, each covering roughly 64 square degrees. These observations have revealed a plethora of structures, whose exact origin is still not fully understood. However, the multi-tracer analysis of the images has resulted in a clear association between neutral magnetised ISM components (dust and atomic hydrogen gas, HI) and structures seen in Faraday depths (\citealt{zaroubi15, kalberla16, jelic18, bracco20}). \citet{vaneck17} have even proposed that the observed emission originates from neutral regions in the ISM where Faraday depolarisation effects are small \citep[e.g. see][and references therein]{sokoloff98}. There are, however, still many uncertainties and the observed correlation between tracers of the multi-phase ISM is limited to only a few single fields. For example, many discovered structures extend outside a single field of view used in a particular study \citep{jelic15}, which again limits the interpretation and calls for analyses over larger areas of the sky.

This endeavour to significantly expand the coverage was {made} by \citet{vaneck19} who used LoTSS preliminary data. They performed Faraday tomography of 568 square degrees in the region of the Hobby-Eberly Telescope Dark Energy Experiment (HETDEX, \citealt{hill08}) spring field. In our work, we continue this effort by expanding the area of analysis even further. Our mosaic covers an area roughly 5.5 times larger than the one in \citet{vaneck19}, making it the largest mosaic of the low-radio frequency polarised emission in the northern sky to date. Moreover, we are using data from LoTSS Data Release 2 \citep[DR2, ][]{shimwell22}, which were calibrated with a more comprehensive procedure than the one used for the preliminary data release \citep{shimwell17}. This resulted in better quality images that, for example, suffer from artefacts around bright sources less and have a lower amount of instrumental polarisation.

The goal of this paper is to present the first LoTSS-DR2 mosaic and highlight some of the main structures we see in this large sky area. While the large mosaic area allows us to gain better insight {into the} origin of observed structures, this paper serves as a first introduction to the LoTSS mosaics and as an in-depth analysis of the different structures found which will be a topic of follow-up papers. Here, we focus on a statistical analysis of the whole mosaic in an effort to shed light on the LOS structure of the diffuse ISM. We achieve this by comparing our Faraday tomographic results with results based on observations at higher frequencies (1.4 GHz) by \citet{dickey19} and with the total Galactic RM map produced by \citet{hutschenreuter22}. The comparison was carried out through statistical Faraday moments, a statistical tool, which was first applied to Faraday spectra by  \citet{dickey19}, that makes it possible to analyse the complexity of Faraday spectra by taking multiple Faraday components weighted by their intensity 
into account. {We used the moments in addition to simple inspection of} the peak intensity and the corresponding depth in the Faraday cube \citep[e.g.][]{jelic15, vaneck17}.

The paper is organised as follows. In Sect. \ref{sec:data}, we describe the LoTSS-DR2 data used in the analyses, as well as the procedure we used to create the mosaic in Faraday depth. The resulting mosaics in equatorial (RA, Dec) and Galactic coordinates ($l$, $b$) are presented in Sect. \ref{sec:mosaic}, together with the results of Faraday tomography highlighting the polarised emission associated with Loop III. Faraday moments of the LoTSS-DR2 data are presented in Sect. \ref{sec:moments}. We discuss them and compare them with high-frequency polarisation data \citep[DRAO GMIMS, ][]{dickey19} and the total Galactic RM map \citep{hutschenreuter22} in Sect. \ref{sec:discussion}. The paper finishes with a summary and conclusions presented in Sect. \ref{sec:end}.

\section{Data and processing}\label{sec:data}

In this section, we describe the {LoTSS-DR2} data and the {derived} data products used in this paper. We give an overview of the {RM synthesis} parameters and the produced Faraday cubes. Finally, we outline the procedure of combining the Faraday cubes into the mosaic cube.

\begin{figure}
   \centering
   \includegraphics[width=\hsize]{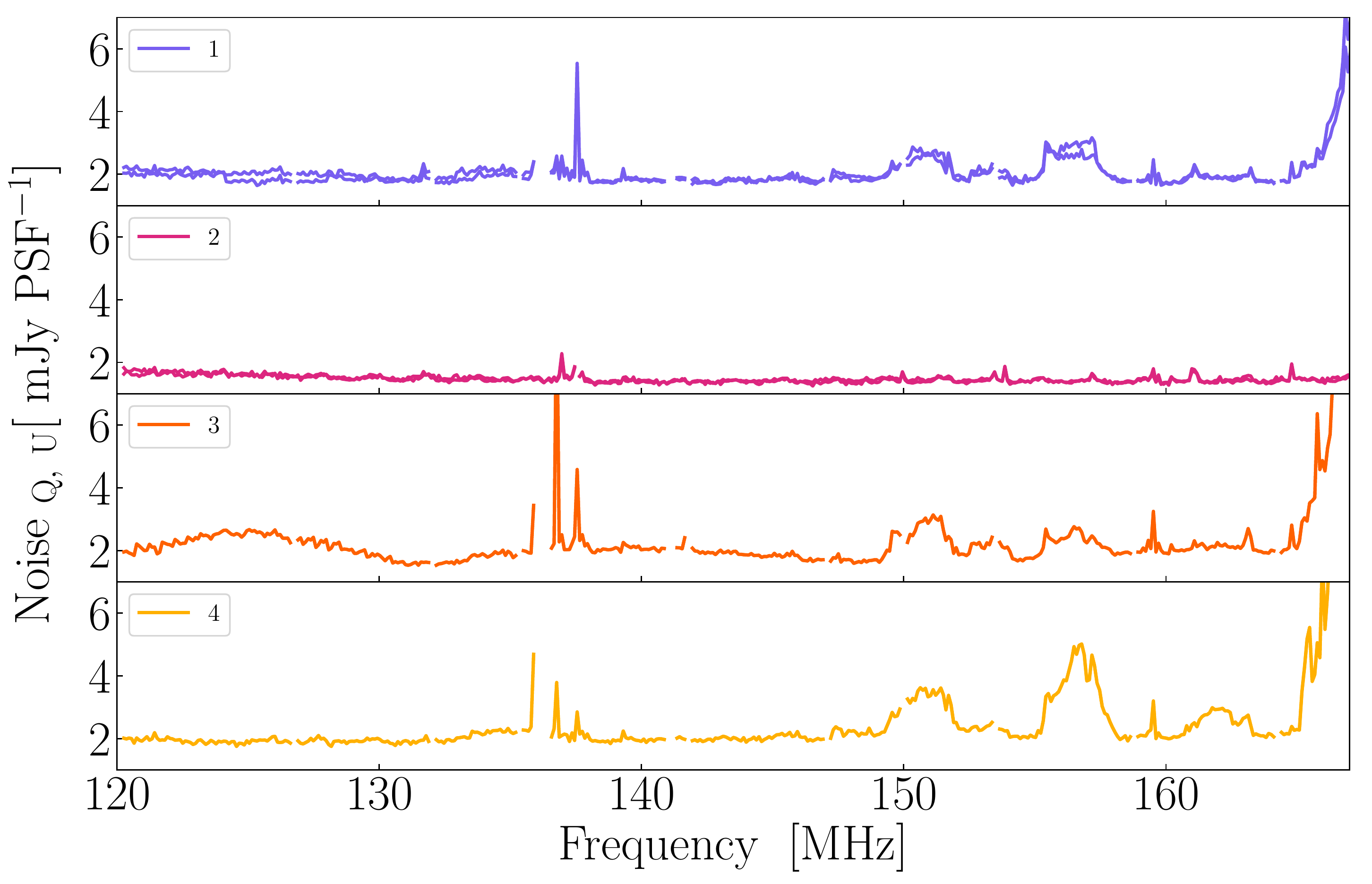}
      \caption{Noise in {both} $Q$ and $U$ cubes not corrected for a primary beam, as a function of the frequency for four fields marked with respective colours {and numbers} in Fig. \ref{mosaic_coverage}. Some of the missing {channels} or {channels} with excess noise are shared among several fields, while others are specific for a single field. The noise excess in the observations comes from the radio frequency interference. {Noise in $Q$ and $U$ is comparable; therefore, the lines mostly overlap.} }
         \label{rnd_noise}
\end{figure}

\begin{table}[h]
 \caption[]{\label{tab_other_fields}Areas in the LoTSS mosaic that were previously studied.}
\begin{tabular}{lcccc}
 \hline \hline
  Field &
  RA &
  Dec &
  Area &
  Reference \\
   & (deg) & (deg) & ($\mathrm{deg}^2$) &
 \\ \hline
HETDEX        &   228.8 - 161.3 & 45 - 57 & 568 & (4)\\
3C196         &  123.4  & 48.2  & 64 & (2), (3), (5)\\
3C196 - B   &   123.4 & 40.4 & 64 & (5), (6)\\
3C196 - C   &   131.2 &  33.9  & 64 & (5), (6)\\
ELAIS - N1  &    242.8 & 55 & 64 & (1)\\

\hline
\end{tabular}
\tablebib{(1) ~\citet{jelic14}, (2) \citet{jelic15}, (3), \citet{jelic18}, (4) ~\citet{vaneck19};
(5) \citet{bracco20}, (6) \citet{turic21}
\label{LoTSSareas}
}
\end{table}

\begin{table}
\caption{\label{table_res} Frequency range and {RM synthesis} parameters in different observations.
                }
\centering
\begin{tabular}{lccc}
\hline\hline
Frequency range & $\delta \phi$ & $\Delta\Phi_\mathrm{max}$ &Field percentage\\

(MHz) & ($\mathrm{rad \ m^{-2}}$) & ($\mathrm{rad \ m^{-2}}$) & (\%)  \\
\hline
120 - 167           & 1.16    & 0.97  & 92.4 \\
120 - 165           & 1.20    & 0.95  & 6.4  \\
120 - 163           & 1.23    & 0.93  & 0.7  \\
120 - 153           & 1.44    & 0.82  & 0.5  \\
\hline
\end{tabular}
\tablefoot{This table shows the resolution in Faraday depth $\delta \phi$  and the largest scale in Faraday depth to which the observations are sensitive. The values change from field to field due to different frequency ranges in cubes (See Fig. \ref{mosaic_coverage}). Here we included only the 440 fields used to make the final mosaic.}
\end{table}

\begin{figure}
   \centering
   \includegraphics[width=\hsize]{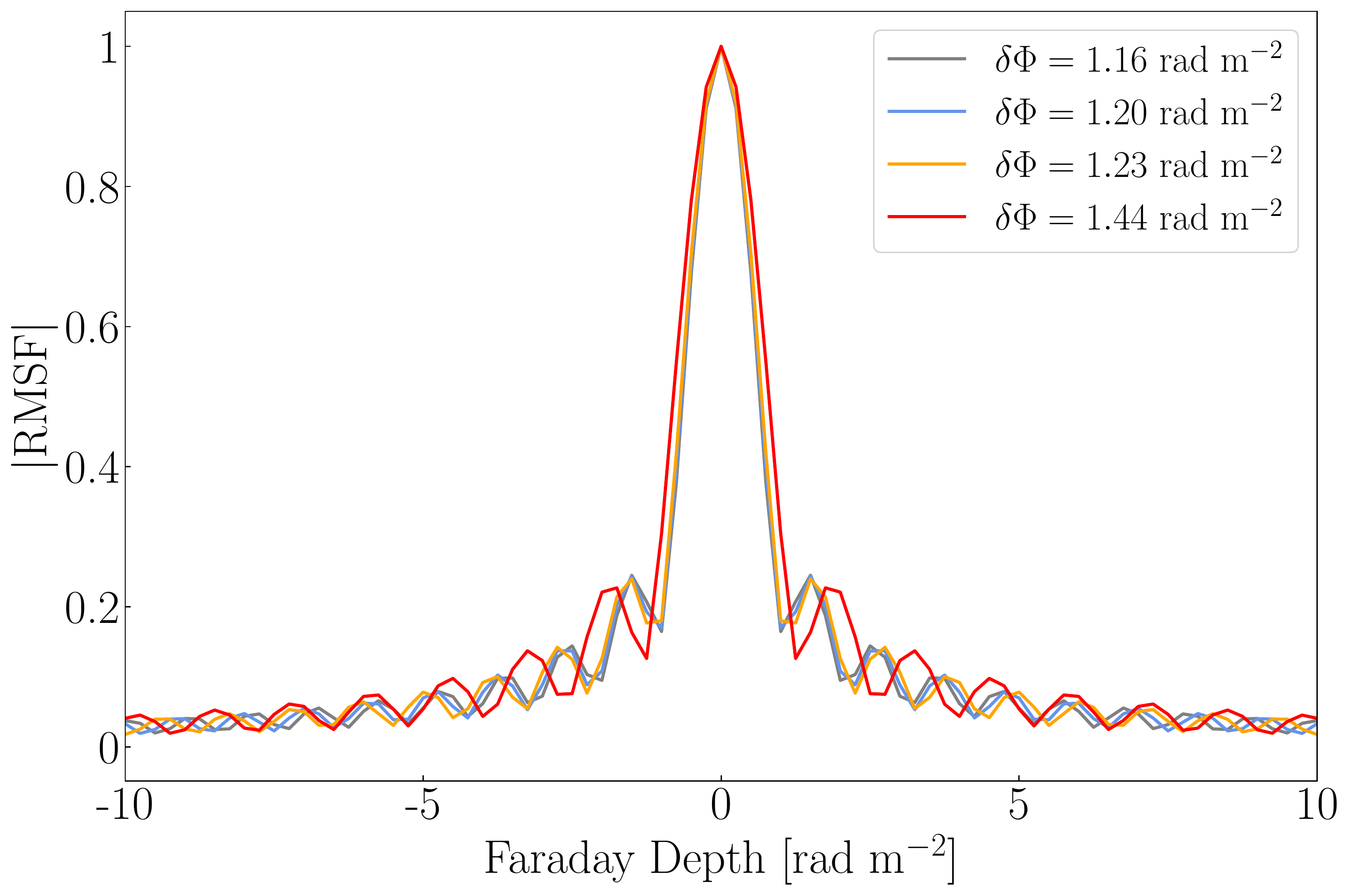}
      \caption{RMSF functions of observations with different upper frequencies (see Table \ref{table_res}). While the {main lobe} of the red function (representing 0.5\% of data) is visibly wider, other functions differ only marginally.
              }
         \label{rmsf}
\end{figure}

\begin{figure}
   \centering
   \includegraphics[width=\hsize]{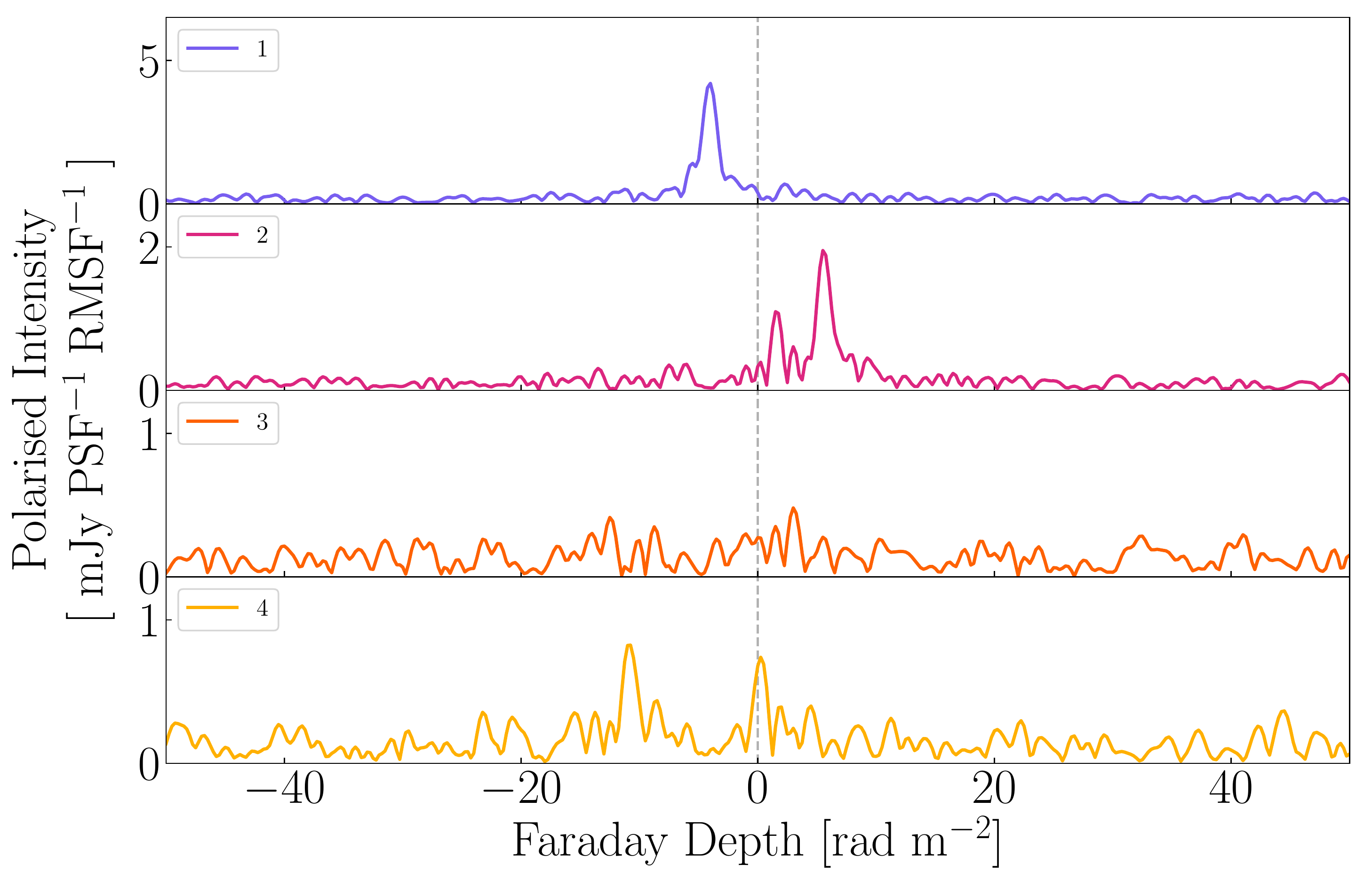}
      \caption{Faraday spectra at four distinct locations in the Faraday cube mosaic. The locations are marked in Fig. \ref{mosaic_coverage} with respective colours {and numbers}. The Faraday spectra shown in this plot are chosen to probe some of the distinct features discussed in Sect. \ref{sec:mosaic}. {Faraday spectra 1 and 4 probe Loop III while Faraday spectra 2 and 3 probe} the zig-zag structure and the very low S/N area in the mosaic, respectively.
              }
         \label{chosen_spectra}
\end{figure}

\begin{figure}
   \centering
   \includegraphics[width=\hsize]{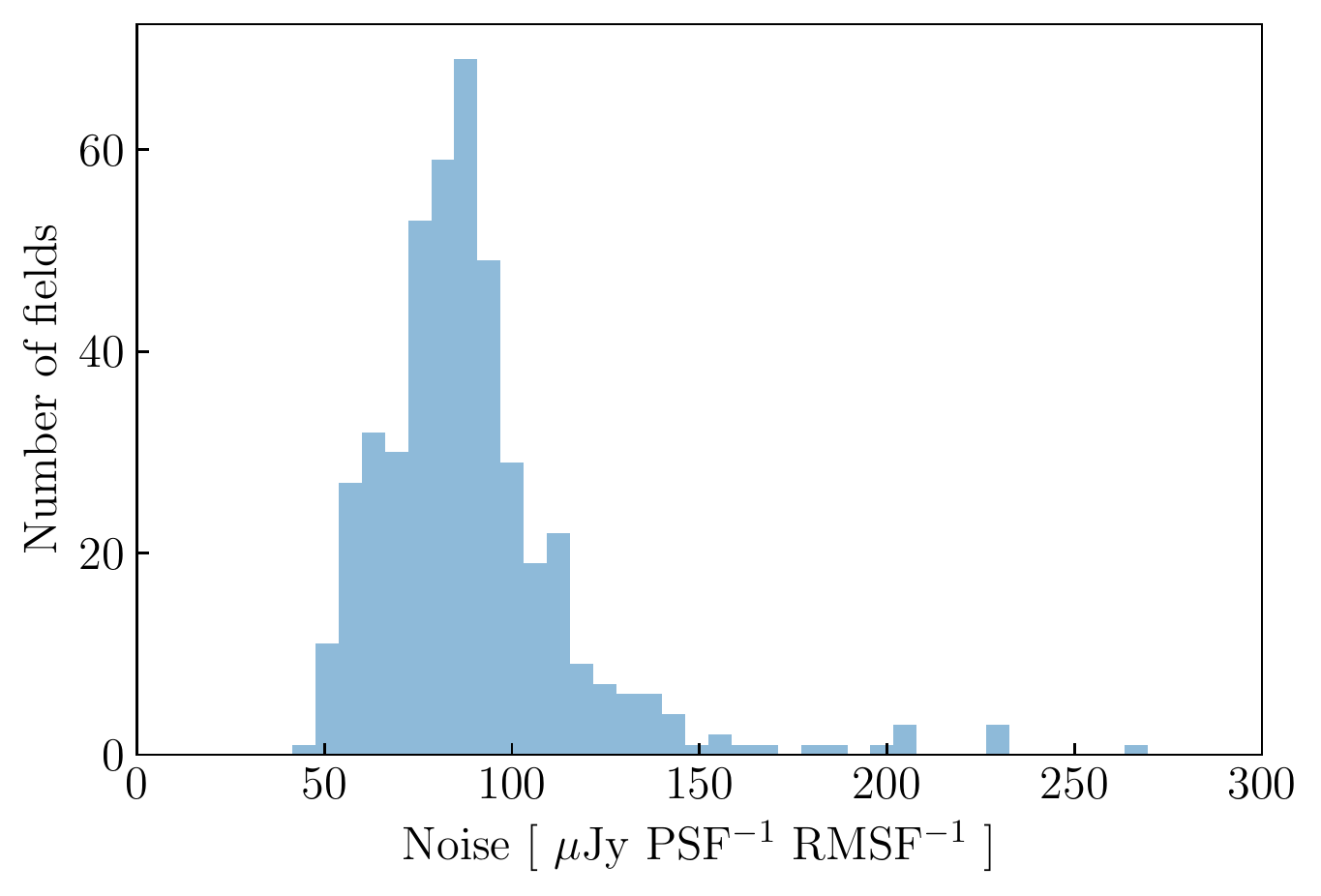}
      \caption{Histogram of noise in each polarised intensity cube, {approximated as a standard deviation multiplied by a factor of $\sqrt{2}$} over each cube at $\Phi = -50~\mathrm{rad~m^{-2}}$. The mean value of this distribution represents the mean value of noise in the produced polarised intensity cubes. 
              }
         \label{noiseP}
\end{figure}

\subsection{{LoTSS-DR2} very low resolution images}
LoTSS \citep{shimwell17, shimwell19} is an ongoing 120~--~168 MHz survey of the northern sky with the LOFAR High Band Antennas  \citep[HBA,][]{vanhaarlem13}. The DR2 of the survey \citep{shimwell22} presents data from 841 individual fields that span a combined area of 5~634 square degrees. This is split between two regions, one centred at $\mathrm{RA= 0^h}$ and another at $\mathrm{RA= 12^h}$. 

The data have been processed in a two-step procedure. The first step was performed with the \texttt{PreFactor pipeline}\footnote{\url{https://github.com/lofar-astron/prefactor}} (\citealt{vanweeren16}, \citealt{degasperin20}), which corrects for direction-independent effects in the data, such as the XX-YY phase offset, {the instrumental time delay associated with clocks}, bandpass, and ionospheric Faraday rotation, before calibrating the data against a sky model derived from the {Tata Institute of Fundamental Research} GMRT Sky Survey \citep[TGSS, ][]{itema17}. This pipeline makes use of software packages such as the \texttt{LOFAR Solution Tool} \citep{degasperin20}, {Default Pre-Processing Pipeline} \citep[\texttt{DPPP}][]{dijkema18}, and \texttt{AOFlagger} \citep{offringa11}. {For efficiency, the pipeline is deployed on the LOFAR archive compute facilities \citep{mechev17,drabent19}.} 

{The second, and more computationally expensive step was to perform direction-dependent calibration and imaging using the \texttt{Direction Dependent Facet (DDF) pipeline}}\footnote{\url{https://github.com/mhardcastle/ddf-pipeline}}, which has been revised for LoTSS-DR2 (and the LoTSS {Deep Fields}) to improve the dynamic range and image fidelity (see \citealt{tasse21} for a full description of the pipeline). This pipeline makes use of the Wirtinger derivative \citep{tasse14,smirnov15} for direction-dependent calibration and \texttt{DDF} \citep{tasse18} to apply these solutions whilst imaging. The processing was primarily performed on the LOFAR UK compute facilities\footnote{\url{https://lofar-uk.org/lucf.html}} in Hertfordshire with a small fraction also done in Hamburg, Leiden, and Bologna. The products from the  DDF pipeline, which are stored in Leiden, include Stokes $I$, $Q$, $U$, and $V$ images at different angular and frequency resolutions, as well as calibration solutions and compressed data sets to allow for re-imaging when necessary.

In this work, we analyse 461 calibrated pointings, covering around 3100 square degrees centred at $\mathrm{RA= 12^h}$. This area covers previously studied areas, as listed in Table \ref{tab_other_fields}. We use the LoTSS-DR2 Stokes $Q$ and $U$ very-low resolution (vlow) images \citep{shimwell22}, which are now publicly available through the LOFAR {Surveys} website\footnote{\url{https://lofar-surveys.org}}. The vlow images have an angular resolution which varies between 5.5 and 4 arcmin across observed frequencies from 120 MHz to 167 MHz (see Table~\ref{table_res}). The frequency resolution of each of the 480 frequency channels is 97.6 kHz. To produce the mosaic, we used both the primary beam corrected and uncorrected images at a common resolution of 5.5 arcmin. 

The data were taken over multiple $\sim$8h observations during different LOFAR observing cycles. The noise is comparable in most of the observations. Figure \ref{rnd_noise} shows the noise in Stokes \textit{$Q$} and $U$ images {not corrected for the primary beam} at different frequencies for {four} different observations. {The noise was calculated as a standard deviation in the corner of the image, where the polarised emission is not present.} Over the observed frequency range, there are some common frequency {channels}, which are affected by broad radio-frequency interference \citep{offringa13}. Each observation also has a few additional low-quality or missing frequency {channels}. To identify those, we used the criterion that the noise in each {channel} should not be higher than 4 $\sigma$ above the mean value of the noise over the full bandwidth, where $\sigma$ is the standard deviation of the noise over the full bandwidth. It is important to note that we identified and flagged bad {channels} in each observation {independently} to avoid flagging a large amount of good quality {channels}. 

\subsection{RM synthesis}
To produce the Faraday cubes based on the LOFAR observations, we performed RM synthesis using \texttt{pyrmsynth-lite}\footnote{\url{https://github.com/sabourke/pyrmsynth_lite}}. We did not weigh the data during this process as the noise is comparable across the selected frequency {channels}. The final cubes span a Faraday depth range from -50 $\mathrm{rad \ m^{-2}}$ to +50 $\mathrm{rad \ m^{-2}}$ in steps of 0.25 $\mathrm{rad \ m^{-2}}$. Faraday spectra were not deconvolved. The sidelobes of the rotation measure spread function (RMSF) are lower than 20\%  and the signal-to-noise ratio (S/N) is, in general, so low that deconvolution would have a very little effect (see Fig.~\ref{rmsf}). Four examples of Faraday spectra found in different areas of the mosaic are shown in Fig. \ref{chosen_spectra}.

The resolution in Faraday depth, defined by the width of the RMSF function, is $\delta\Phi=1.16~{\rm rad~m^{-2}}$ for 92.4\% of the observations (see Table~\ref{table_res}). Some observations have marginally poorer resolution due to a slightly lower upper frequency in the observed range (see Sect. \ref{sec:intro}and Fig. \ref{rmsf}). Similarly, there is also a small difference in the largest Faraday structure that can be resolved in some observations ($\Delta\Phi_\mathrm{max}$, see Table~\ref{table_res}). In all our observations, only Faraday thin structures can be detected, or the edges of Faraday thick structures (as described in \citealt{brentjens05, vaneck17}), as the resolution in Faraday depth is comparable to the largest observable Faraday scale.

The histogram in Fig.~\ref{noiseP} shows the distribution of the noise estimated from the polarised intensity Faraday cube for each observation. The noise was calculated as the standard deviation of the polarised intensity at $\Phi = -50~{\rm rad \ m^{-2}}$ {multiplied by a factor of $\sqrt{2}$}. {This factor comes from the noise in polarised intensity being distributed according to the Rician distribution and it is equivalent to the normally distributed noise in Q and U \citep[e.g.][]{brentjens05, hales12}. The image at this Faraday depth is dominated by noise and the noise value obtained from it is consistent with noise at other noise-dominated Faraday depths.} The distribution has a mean value of {117} $\mathrm{\mu Jy~PSF^{-1}~RMSF^{-1}}$, which is lower by roughly a square root of the number of frequency channels compared to the noise in the Stokes $Q$ and $U$ images given at different frequencies ($\sim1.8~\mathrm{mJy~PSF^{-1}}$), as expected. Some observations have much higher noise values and show systematic artefacts. The 21 cubes (out of the total 461) in which the noise exceeds {200} $\mathrm{\mu Jy~PSF^{-1}~RMSF^{-1}}$ are excluded from further analyses. The total number of cubes used in Sect. \ref{sec:mosaic} to construct the mosaic is 440.

\subsection{Mosaicing}
We used \texttt{Montage}\footnote{\url{http://montage.ipac.caltech.edu}}, an {Astronomical Image Mosaic Engine}, to re-project and combine the Faraday cubes from different observations into a single mosaic cube. This software package has the functionality to deal with different projections while conserving the flux.  When re-projecting on curved grids (in our case from a gnomonic sinusoidal projection to a curved Aitoff projection), pixel areas change and conservation of the flux needs to be taken into account by redistributing the flux from one grid to another. We used \texttt{Montage} modules \texttt{mProjExec} and \texttt{mAdd} for this purpose. The flux {is} redistributed by weighing each input pixel by the sky area of overlap with the output pixel. When combining multiple overlapping pixels, it assigns the area-weighted average of the pixels. 

Additionally, we introduced a weighting scheme to account for the following: (i) variations of the noise across Faraday cubes of different observations and (ii) overlaps between them. We weighed each observation by its noise and by the primary beam attenuation factor (pb):
\begin{equation}
w_{\rm obs} = \frac{1}{\mathrm{\sigma_{\rm obs}}^2}*\frac{1}{\mathrm{pb^2}},
\end{equation}
where $\sigma_{\rm obs}$ is the standard deviation of the noise in each primary beam uncorrected Faraday cube, calculated at a Faraday depth of $-50~{\rm rad~m^{-2}}$. The primary beam was assessed by dividing the primary beam corrected cube by the uncorrected cube. 

Being a square root quantity, the noise in the polarised intensity Faraday cube does not have a normal distribution as the noise in Stokes $Q$ and $U$ Faraday cubes. Therefore, the mosaicing was done in Stokes $Q$ and $U$ after which we created the polarised intensity mosaic cube, as $P=\sqrt{Q^2+U^2}$. 

In practice, we first created a weighted mosaic of primary beam corrected images (${\rm img_{pbcorr}^{Q,U}}$) in Stokes $Q$ and $U$ at each Faraday depth $\Phi$:
\begin{equation}
        \mathrm{Mosaic^{Q,U}(\Phi)} = \mathrm{\frac{mosaic (img_{pbcorr}^{Q,U}(\Phi) * w_{\rm obs})}{mosaic (w_{\rm obs})}}.
\end{equation}
The created 2D mosaics were then combined into a mosaic Faraday cube.

\begin{figure*}
   \centering
   \includegraphics[width=\textwidth]{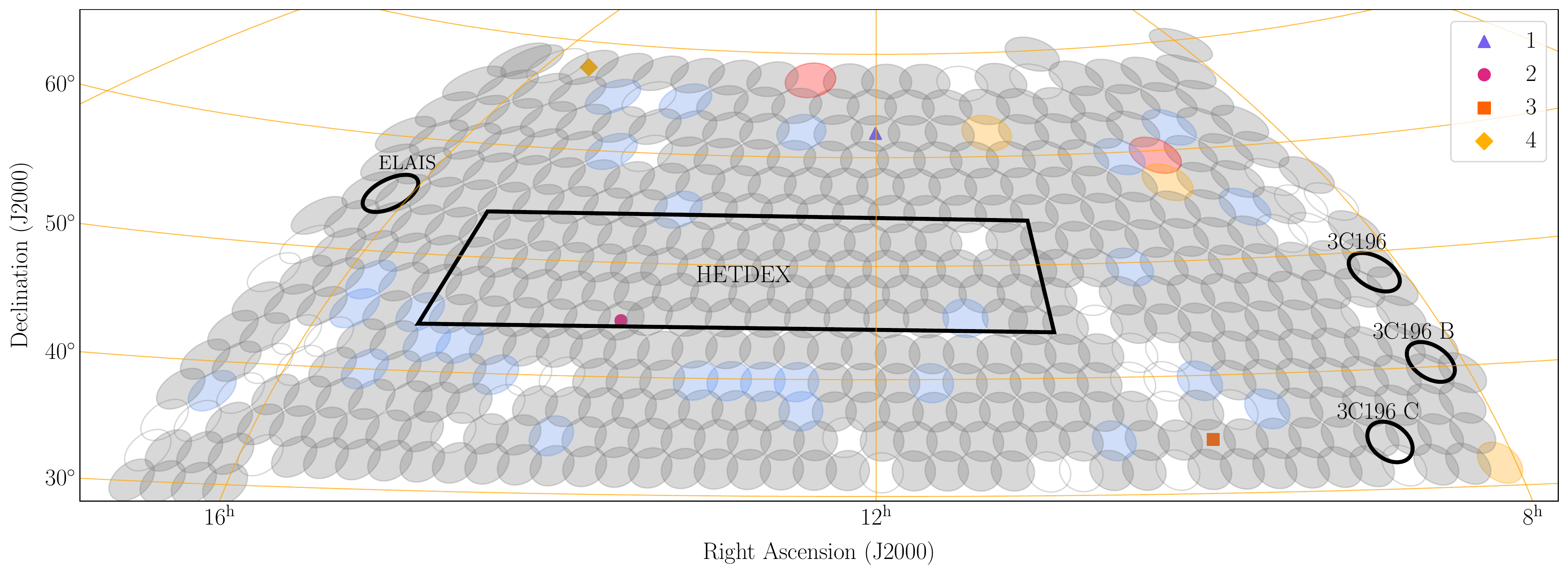}
      \caption{Mosaic coverage shown in equatorial coordinates. Radii of the circles correspond to the largest primary beam FWHM of the LoTSS survey ($4.75^\circ$). Red, orange, blue, and grey circles mark Faraday space resolution of  1.44, 1.23, 1.20, and 1.16 $\mathrm{rad \ m^{-2}}$, respectively (see Table \ref{table_res}). Areas previously studied by different authors are outlined in black (see Table \ref{tab_other_fields} for more information). {Symbols in the legend} mark fields whose noise and spectra are shown in Figs. \ref{rnd_noise} and \ref{chosen_spectra}, {where they are drawn with a corresponding colour and marked with numbers 1 through 4}. Empty circles outlined in grey mark bad quality or missing observations.}\label{mosaic_coverage}
\end{figure*}

\section{The LoTSS mosaic Faraday cube} \label{sec:mosaic}
The final mosaic spans an area from $\mathrm{8^h00^m}$ to $\mathrm{16^h40^m}$ in right ascension and from $\mathrm{30^\circ}$ to $\mathrm{70^\circ}$ in declination. Figure \ref{mosaic_coverage} gives an overview of all LoTSS observations in this area in equatorial coordinates. Filled-in circles correspond to the location of 440 observations used in the final mosaic.  Different colours illustrate small variations in the Faraday depth resolution in different fields. Empty circles are observations that were removed due to high noise or simply missing observations.

The central positions of observations are typically separated by $2.58^\circ$. Each pointing has six nearest neighbours within $2.80^\circ$. The primary beam full width half maximum (FWHM) of each observation ranges from $3.40^\circ$ to $4.75^\circ$ over the observed frequencies \citep{shimwell19}, so that the LoTSS survey has a very good coverage. Thus, the removal of 21 observations from the final mosaic cube {did not} affect the coverage greatly, except in places where several neighbouring pointings were removed. In these areas none of the surrounding primary beams overlap within their FWHM, so we excluded them from our further analysis. 

\begin{figure}
   \centering
   \includegraphics[width=\hsize]{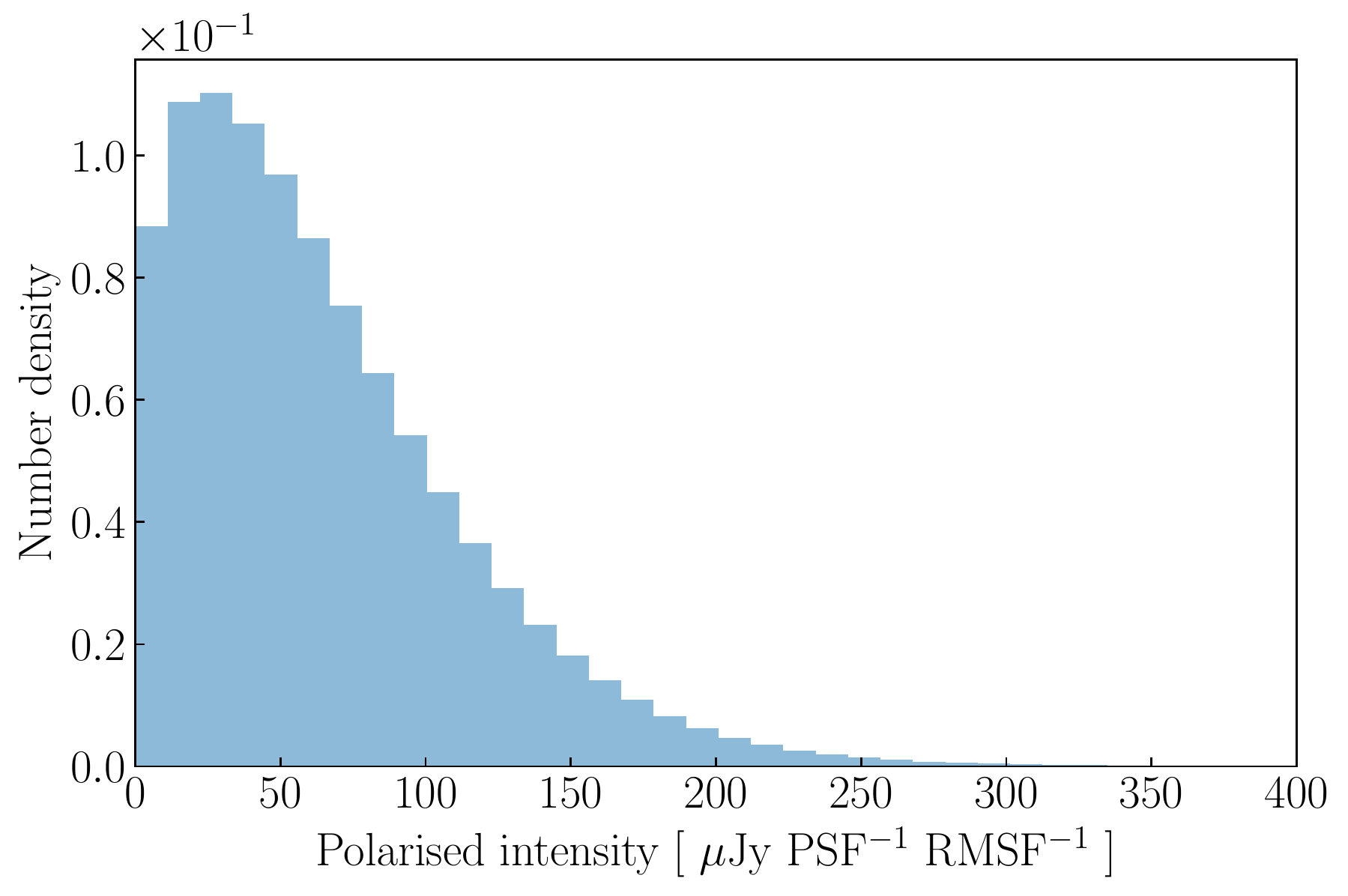}
      \caption{Histogram of polarised intensity of the mosaic at Faraday depth $\Phi = -50 \ \mathrm{rad \ m^{-2}}$ (Fig. \ref{mosaic_noise}). The width of this distribution, {approximated by the standard deviation multiplied by $\sqrt{2}$, gives an estimate of the noise level in the mosaic, 71 $\rm{\mu Jy \ PSF^{-1} \ RMSF^{-1}}$,} while its mean gives the polarisation bias, 66 $\rm{\mu Jy  \ PSF^{-1} \ RMSF^{-1}}$.}   \label{noise_hist}
\end{figure}

\begin{figure*}
   \centering
   \includegraphics[width=\textwidth]{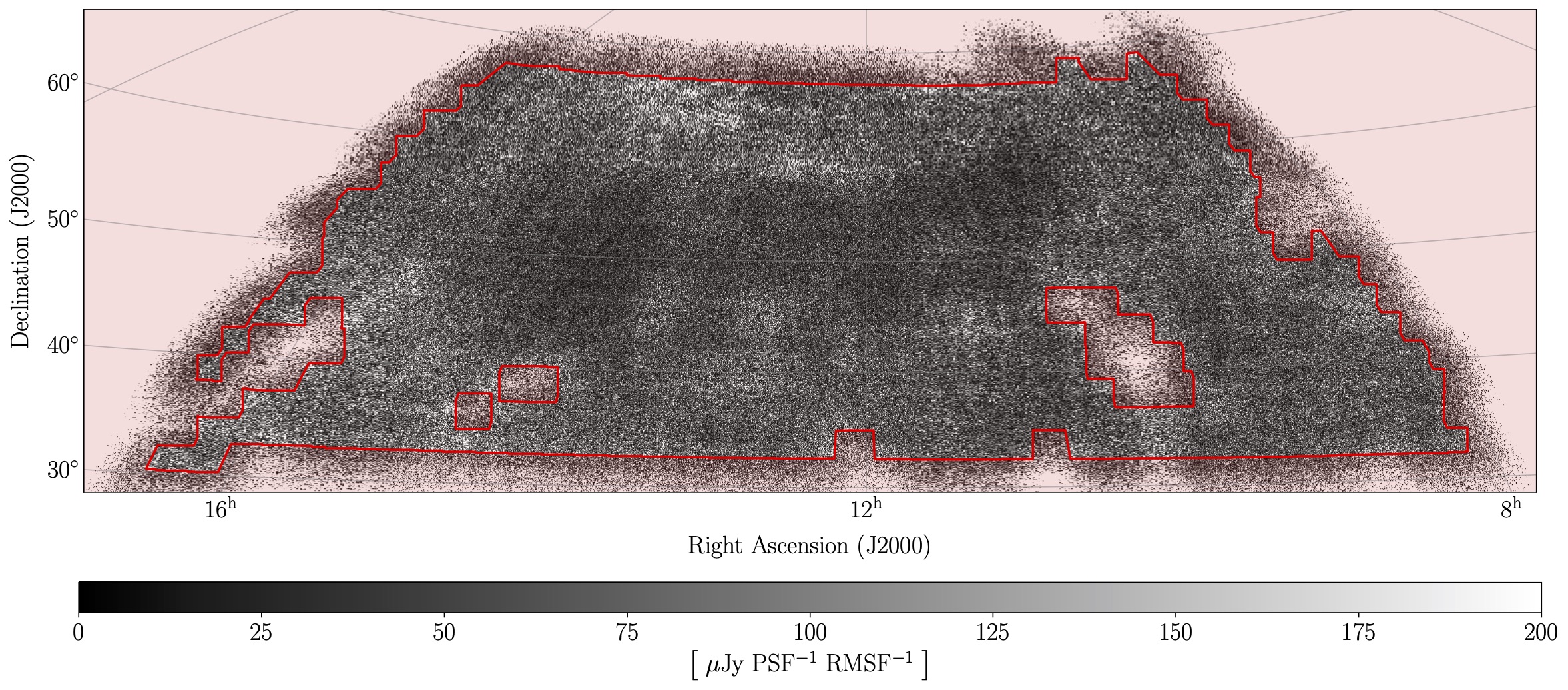}
      \caption{Polarised intensity image of the mosaic at Faraday depth $\Phi = -50 \ \mathrm{rad \ m^{-2}}$, corresponding to the noise level across the mosaic. The red contour marks the mask used in analyses throughout the paper to exclude areas shaded in red. }
         \label{mosaic_noise}
\end{figure*}

The noise across the mosaic is represented by polarised intensity at $\Phi = -50~\mathrm{rad~m^{-2}}$, whose distribution is shown in Fig. \ref{noise_hist}. A typical value for the noise, that is the width of distribution in Fig. \ref{noise_hist}, is $50~{\rm \mu Jy~PSF^{-1}~RMSF^{-1}}$. This was calculated excluding the regions where the noise is boosted by the primary beam (marked with a red contour in Fig. \ref{mosaic_noise}). These regions are mostly at boundaries of the mosaic and in the aforementioned areas with the excluded observations. The mask marked with red contours in Fig. \ref{mosaic_noise} is used for statistical analysis of the data throughout the paper. There is a positive polarised intensity bias of $66~{\rm \mu Jy~PSF^{-1}~RMSF^{-1}}$.  We did not correct the data for this bias as it did not affect the qualitative analysis that is the focus of this paper.

\begin{figure*}[ht]
   \centering
   \includegraphics[width=\textwidth]{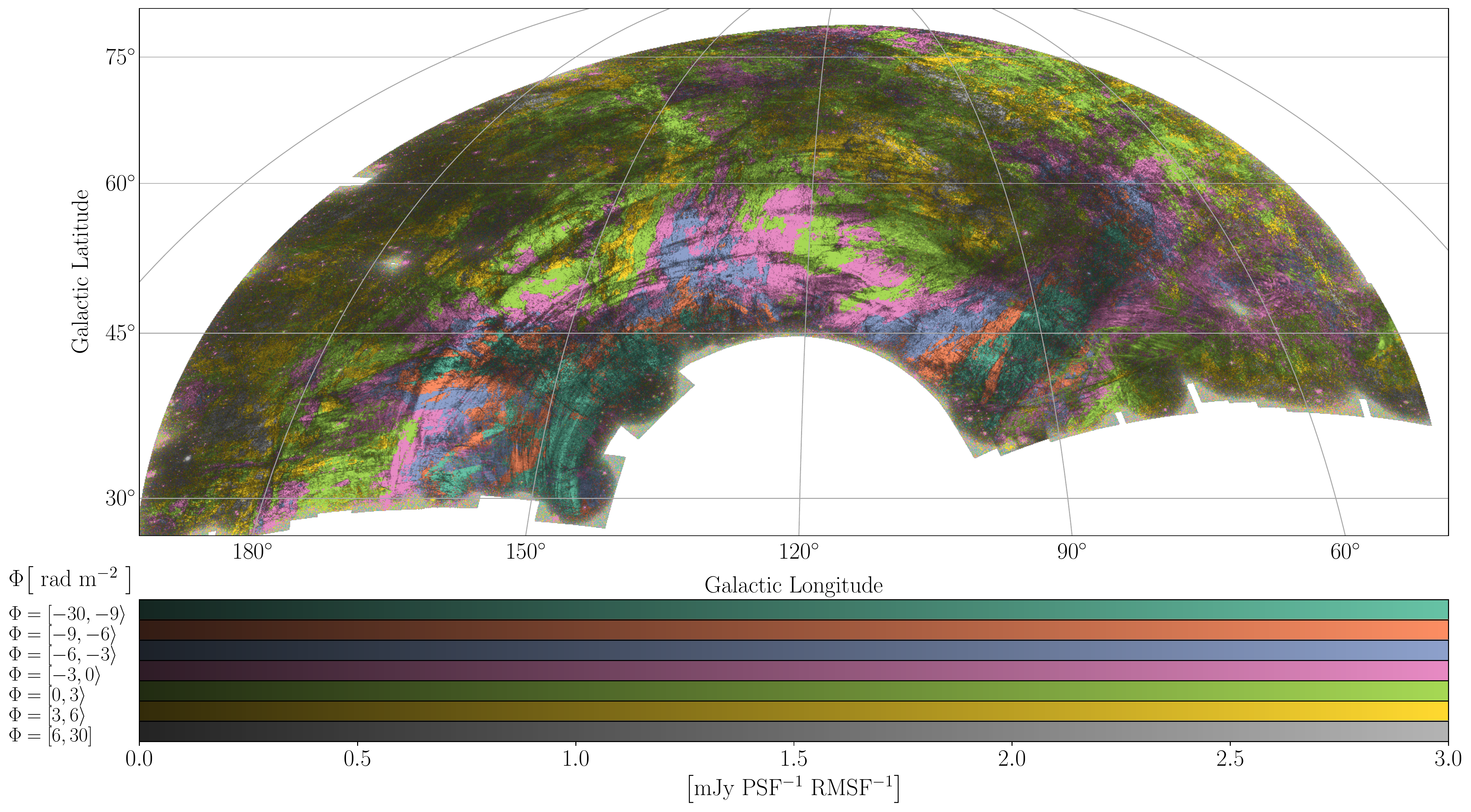}
      \caption{Maximum polarised intensity per Faraday depth range in the LoTSS mosaic {in Galactic coordinates (Mollweide projection)}, coloured based on the Faraday depth range at which the peak of the emission is found. {The intensity of 1 $\rm{mJy~PSF^{-1}~RMSF^{-1}}$ corresponds to  $\sim 0.5~\rm{K~RMSF^{-1}}$.}
                    }
         \label{colored_gal}
\end{figure*}

\begin{figure*}
   \centering
   \includegraphics[width=\textwidth]{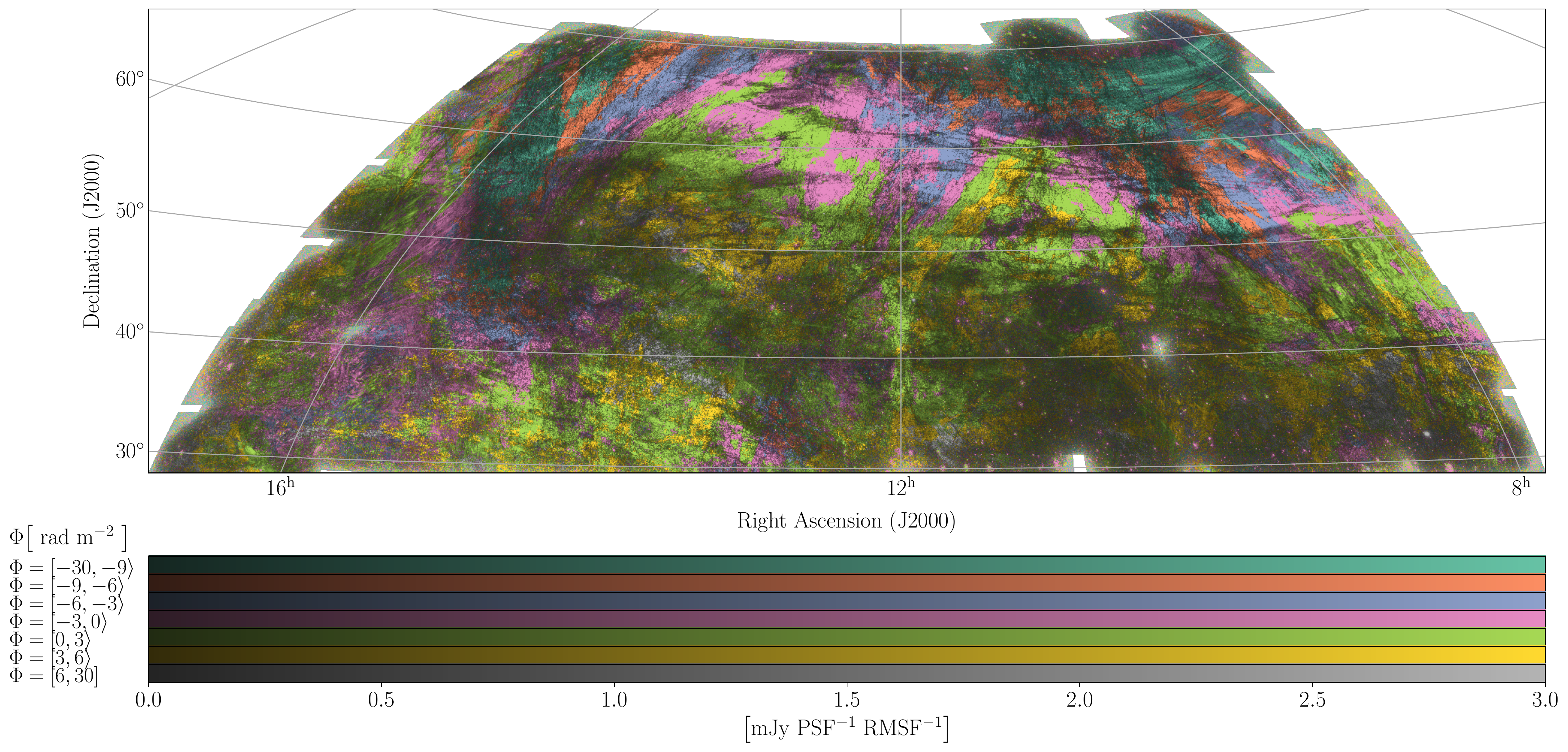}
      \caption{Maximum polarised intensity per Faraday depth range in the LoTSS mosaic in equatorial coordinates, coloured based on the Faraday depth range at which the peak of the emission is found. The intensity of 1 $\rm{mJy~PSF^{-1}~RMSF^{-1}}$ corresponds to $\sim 0.5~\rm{K~RMSF^{-1}}$. 
                  }
         \label{mosaic_colored}
\end{figure*}

\begin{figure*}
   \centering
   \includegraphics[width=\textwidth]{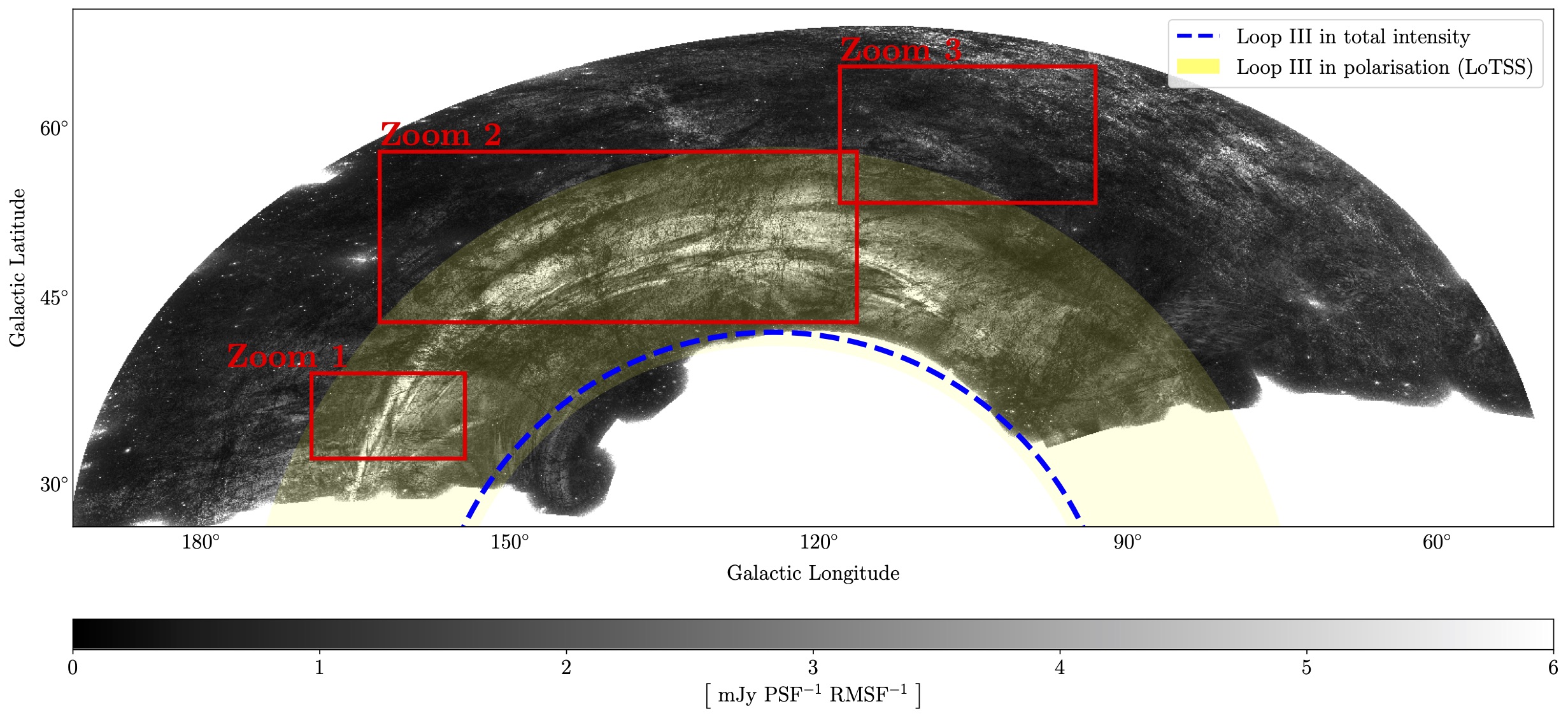}
      \caption{Maximum polarised intensity image {given in Galactic coordinates (Mollweide projection) with some of the interesting features outlined}. The area shaded in yellow is the area in which we see Loop III in LoTSS data. The blue dashed line represents the location of Loop III in total intensity as seen by \citet{berkhuijsen1971a}. Red rectangles mark areas that are magnified in Appendix \ref{app:zoom}, Figs. \ref{zoom1}, \ref{zoom2}, \ref{zoom3} .}
         \label{mosaic_map}
\end{figure*}

Figures \ref{colored_gal} and \ref{mosaic_colored}  summarise the overall results of Faraday tomography in the mosaic area. They give the maximum intensity of Faraday spectra for each pixel in the mosaic, presented in equatorial and Galactic coordinates, respectively. Each presented projection has some advantages for the analysis of the observed structures. Different colours in Figs. \ref{colored_gal} and \ref{mosaic_colored} illustrate the emission observed at different Faraday depth ranges and the colour brightness indicates intensity of the emission. We chose the Faraday depth range from -30 to 30 $\mathrm{rad \ m^{-2}}$ as most of the emission is found there (see Appendix \ref{app:mosaic_slices}). The {colour bins were} chosen to best present {large} structures {visible in the mosaic since the full complexity of a 3D Faraday cube is difficult to represent in a single 2D image. The binning is finer at negative Faraday depths, where we see more emission peaks than at positive Faraday depths. This choice of binning is not necessarily optimal for every region of the mosaic; some local features might be better represented with a different binning.} In the following subsections, we further discuss the {emission features}. 

\subsection{Polarised emission associated with Loop III} \label{subsec:3.1}
The mosaic viewed in Galactic coordinates (Figs.~\ref{colored_gal}, \ref{mosaic_map} ) reveals a striking loop-like structure dominating a large area in the bottom part of the mosaic. This structure is in close proximity and follows the shape of the radio Loop III, centred at $l=124^\circ$ and $b=15.5^\circ$ in total intensity \citep{berkhuijsen1971a}. The polarised emission of the loop is also seen at higher radio frequencies \citep[][Peel et. al, in prep.]{vidal15, planck16}; however, it is located several degrees {in latitude higher than the} total intensity emission. In the LoTSS data, at low frequencies, the loop-like emission mostly overlaps with the Loop III polarised emission at higher frequencies; however, its emission peaks at latitudes 5 to 10 degrees higher and extends {to latitudes} several degrees further up. We therefore associate the loop-like structure observed in the bottom part of the mosaic with the polarised emission of radio Loop III. The distance to Loop III was estimated to be between 150 pc and 250 pc \citep{spoelstra72, lallement03, kun07}, which is in agreement with the observed large angular size of tens of degrees across the sky.

Towards the edge of Loop III, we found continuous depolarisation canals following the loop curvature and stretching over tens of degrees. Depolarisation canals are narrow and {extended} regions in the image where {the emission is significantly reduced}, which we mostly associate with beam depolarisation in areas showing discontinuity in polarisation angles \citep{haverkorn04, jelic15}. There are also depolarisation canals which are perpendicular to the direction of the loop. They are less prominent, shorter (several degrees in length), and more densely distributed compared to the aforementioned depolarisation canals. The observed structures are evidently not completely disordered. Their morphology suggests connection with a large-scale shell, which was possibly formed by one or more supernovae at the centre of the loop \citep{berkhuijsen1971a, kun07}. One has to bear in mind that the shortest baselines used in LoTSS observations (100 m) allow our mosaic to be sensitive to emission on an angular scale up to $\sim 1^{\circ}$ \citep{shimwell17}. Therefore{, in the mosaic we might be} missing the contribution from angular scales larger than $\sim 1^{\circ}$. {Nevertheless, in some regions of the mosaic, we see structures spanning over several single field observations. Continuous organisation of these small-scale structures points to a common underlying large-scale morphology.}

At different Faraday depths, polarised emission associated with the loop seems to `travel' from the loop centre outwards, creating a gradient from $-30~{\rm rad~m^{-2}}$ at lower Galactic latitudes to $+6~{\rm rad \ m^{-2}}$ at higher Galactic latitudes (see Faraday depth mosaic movie and Fig.~\ref{colored_gal}). The gradient is characterised by changes of $\sim 0.5 - 1~{\rm rad~m^{-2}~deg^{-1}}$ {mostly in the direction} parallel to the loop edge. {However, due to the complex morphology of this emission, the gradient direction is perpendicular to the loop in some areas (e.g. westernmost part of the loop, around $l=100^{\circ}$). A more detailed analysis of the observed Loop III emission gradient is out of the scope of this paper.} At Faraday depths higher than $+6~{\rm rad~m^{-2}}$, there is almost no emission associated with the loop. The mean polarised intensity of emission in the loop area is $2~{\rm mJy~PSF^{-1}~RMSF^{-1}}$. Although not recognised as such, a small part of the loop was already seen in the LoTSS-PDR mosaic that covered the HETDEX spring field region and was analysed by \citet{vaneck19}.

\subsection{Other morphological features of polarised emission}\label{subsec:3.2}
Due to projection effects present in Galactic coordinates, an equatorial projection (Fig. \ref{mosaic_colored}) is used to better highlight {other} structures in the area surrounding Loop III. While most of the emission unrelated to the loop is patchy and faint, one structure is clearly present at positive Faraday depths in the south-eastern area of the mosaic. This zig-zag structure is located between RA $13^h00^m$ and $15^h00^m$ and Dec $30^\circ$ and $55^\circ$. {The emission associated with this structure first appears at around +2 $\mathrm{rad \ m^{-2}}$ and then builds up towards higher Dec creating a gradient up to $\sim$+10 $\mathrm{rad \ m^{-2}}$. The structure is best seen at +8 $\mathrm{rad \ m^{-2}}$} {(see Appendices \ref{app:zoom}, \ref{app:mosaic_slices}, and the Faraday cube movie)}. The mean polarised intensity of this structure is 1.2 ${\rm mJy~PSF^{-1}~RMSF^{-1}}$. 

The emission coming from the zig-zag structure is very hard to disentangle from Loop III emission in the Faraday depth range from 0 to 6 $\mathrm{rad \ m^{-2}}$. It could be a separate structure as it shows no clear morphological connection to the loop at lower or higher Faraday depths, and it has a distinctive configuration at Faraday depths higher than roughly 6.5 $\mathrm{rad \ m^{-2}}$. However, we cannot exclude the possibility that it is connected to Loop III emission. The zig-zag structure seems to be connected to the south-west filament from the LoTSS-PDR polarisation images of the HETDEX field \citep{vaneck19}, where it is not fully visible, in part because it {extends} outside of the field of view, but mainly because it is in the area of a big artefact around the source 3C~295. Due to significant improvements \citep{tasse21, shimwell22} in the data calibration since the preliminary data release \citep{shimwell17}, the artefact is no longer present in our data. We discuss {the zig-zag} structure again in Section~\ref{sec:moments}.

The central part of the south-western quarter of the mosaic is characterised by consistently low polarised intensity of{} 0.8 ${\rm mJy~PSF^{-1}~RMSF^{-1}}$, on average, which is two times lower than the mosaic average of 1.6 ${\rm mJy~PSF^{-1}~RMSF^{-1}}$. Part of this area (west from $9^h$) is in the vicinity of the 3C 196 radio source and it has been previously studied by \citet{jelic15, zaroubi15, jelic18, bracco20, turic21}. The LoTSS mosaic shows that this westernmost region, as well as the easternmost region of the mosaic, has many more depolarisation canals that are seen to extend well beyond the fields in which they were observed up until now. The characteristics of the depolarisation canals found in the mosaic will be investigated in future work.

\section{Faraday moments}\label{sec:moments}
Faraday moments are a useful alternative tool for statistical analysis of Faraday tomographic cubes, giving us information similar to the maximum polarised intensity and the {corresponding} Faraday depth in images presented in the previous subsection (Figs. \ref{colored_gal}, \ref{mosaic_colored}). Both of these methods are a way to represent 3D spectral information from a Faraday cube in 2D images. The advantage of using statistical moments, over simply finding the maximum peak in the spectra, is the ability to take the complexity of spectra containing more than one Faraday structure  into account, as is discussed in the appendix of \citet{dickey19}. Following \citet{dickey19}, we calculated the three moments of the lowest order. 

The zeroth Faraday moment, $M_0$, is the polarised intensity integrated over the full Faraday depth range, given in units of $\mathrm{Jy \ PSF^{-1} \ RMSF^{-1} rad \ m^{-2} }$. The zeroth moment is similar to the maximum polarised intensity image, particularly in areas where Faraday spectra are dominated by a single component. Differences occur in areas where Faraday spectra have multiple components{, which have a} similar polarised intensity. 

The first Faraday moment, $M_1$, is the polarised intensity weighted mean of Faraday spectra in units of $\mathrm{rad \ m^{-2}}$. It gives similar information to the colour in Figs. \ref{colored_gal}, \ref{mosaic_colored}, showing Faraday depth of the maximum polarised intensity. However, it gives a better estimate of the mean Faraday depth, $\Phi_{\mathrm{mean}}$, as it takes multiple components present in complex Faraday spectra into account.

The second Faraday moment, $M_2$, is the intensity weighted variance of Faraday spectra. The square root of the second moment gives the spread of the Faraday spectrum in units of $\mathrm{rad \ m^{-2}}$. This variable quantifies the complexity of Faraday spectra -- low values were obtained for spectra dominated by one component and high values were obtained for multi-component spectra. It can also be interpreted as the width of the Faraday depth range within which {the polarised emission is observed.} 

When calculating Faraday moments, one has to exclude noise-dominated areas in the data, otherwise these areas would give rise to large errors. This is especially true for the second moment image, where adding in quadrature increases the effect of noise. {In our analysis, we use a threshold of $460~\mathrm{\mu Jy \ PSF^{-1} \ RMSF^{-1}}$, which is around six times the typical level of noise in polarised intensity, and we exclude all pixels with lower values.} This condition completely masks regions with Faraday spectra {such as} the one represented by Faraday spectrum 3 in Fig. \ref{chosen_spectra}. We also exclude the masked area shown in Fig.~\ref{mosaic_noise}, where the noise is boosted by the primary beam either at the boundaries of the mosaic or in the regions where we removed bad quality observations. In total, around 40$\%$ of the original data are not used in our Faraday moment analysis. 

\begin{figure*}
   \centering
   \includegraphics[width=0.96\textwidth]{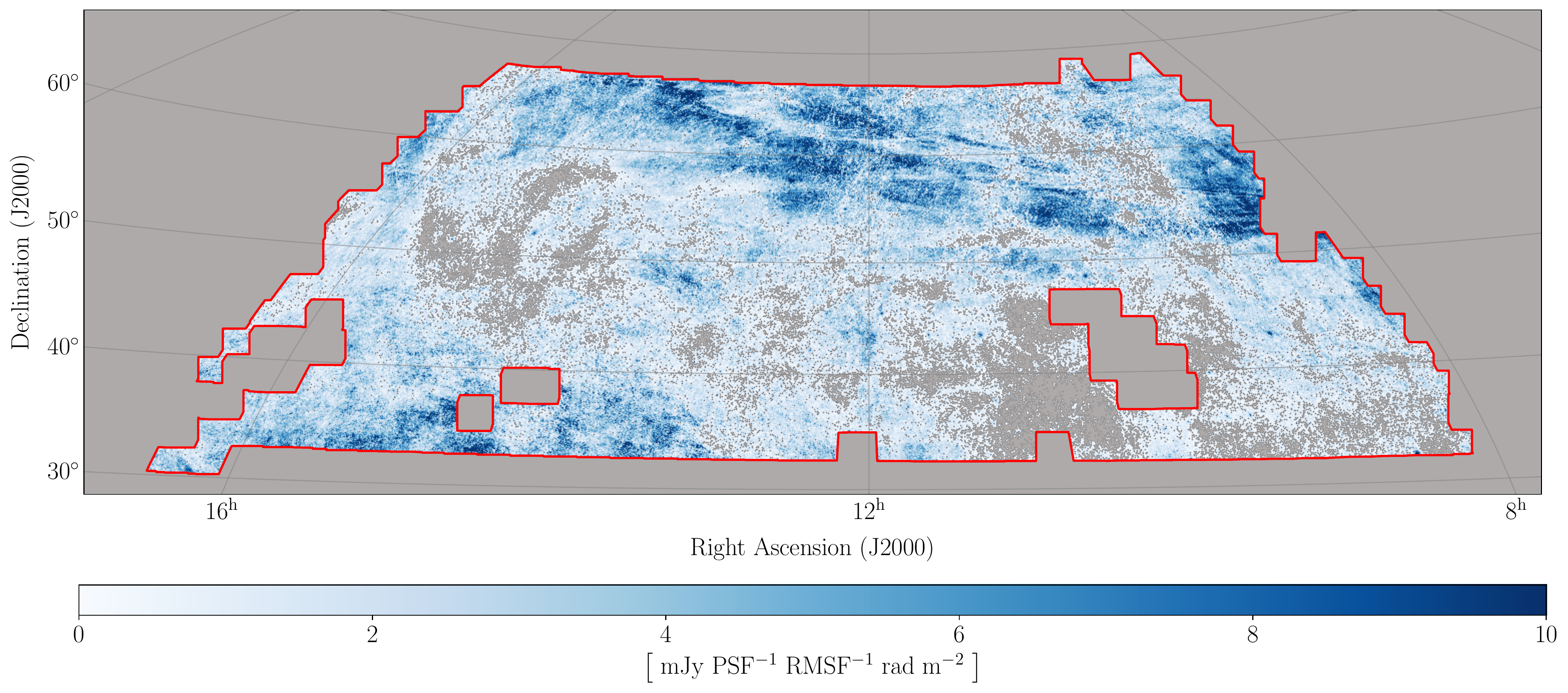}
   \includegraphics[width=0.96\textwidth]{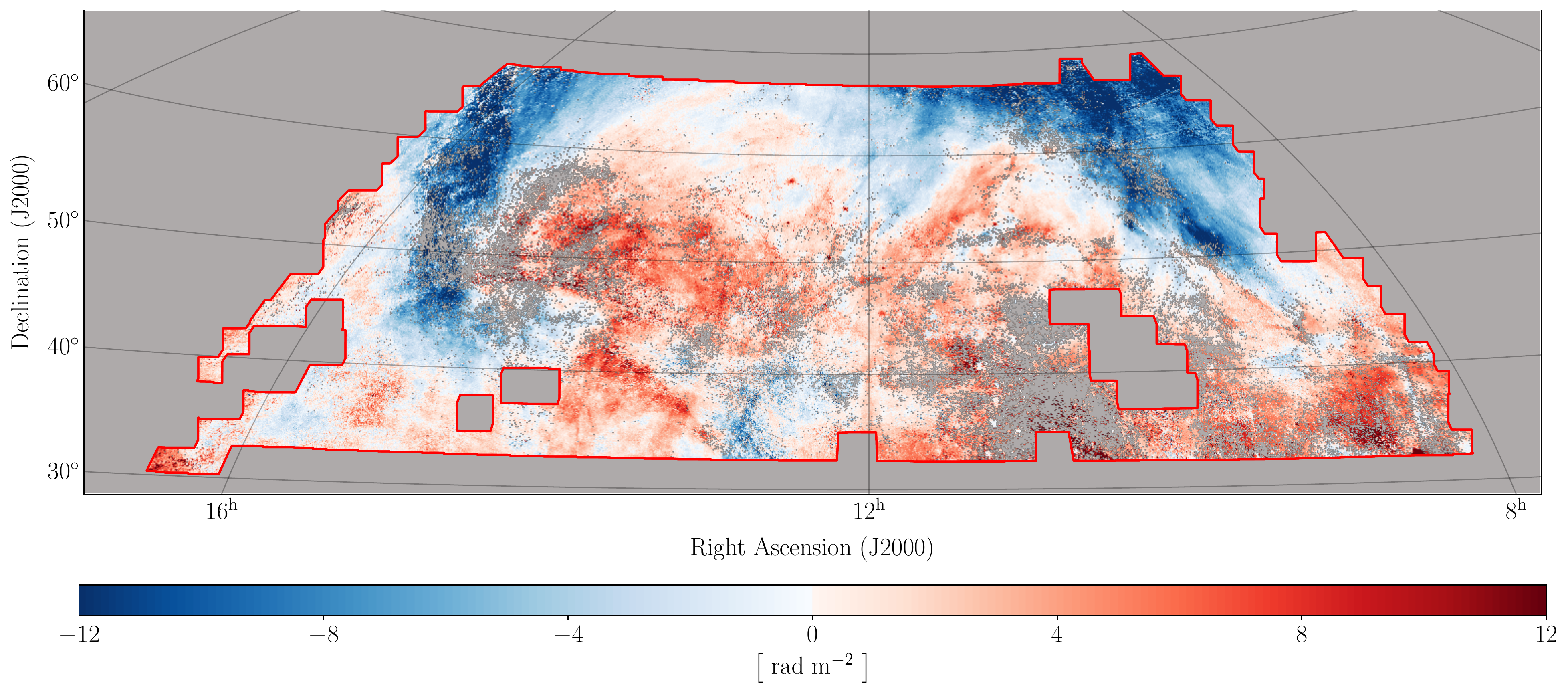}
   \includegraphics[width=0.96\textwidth]{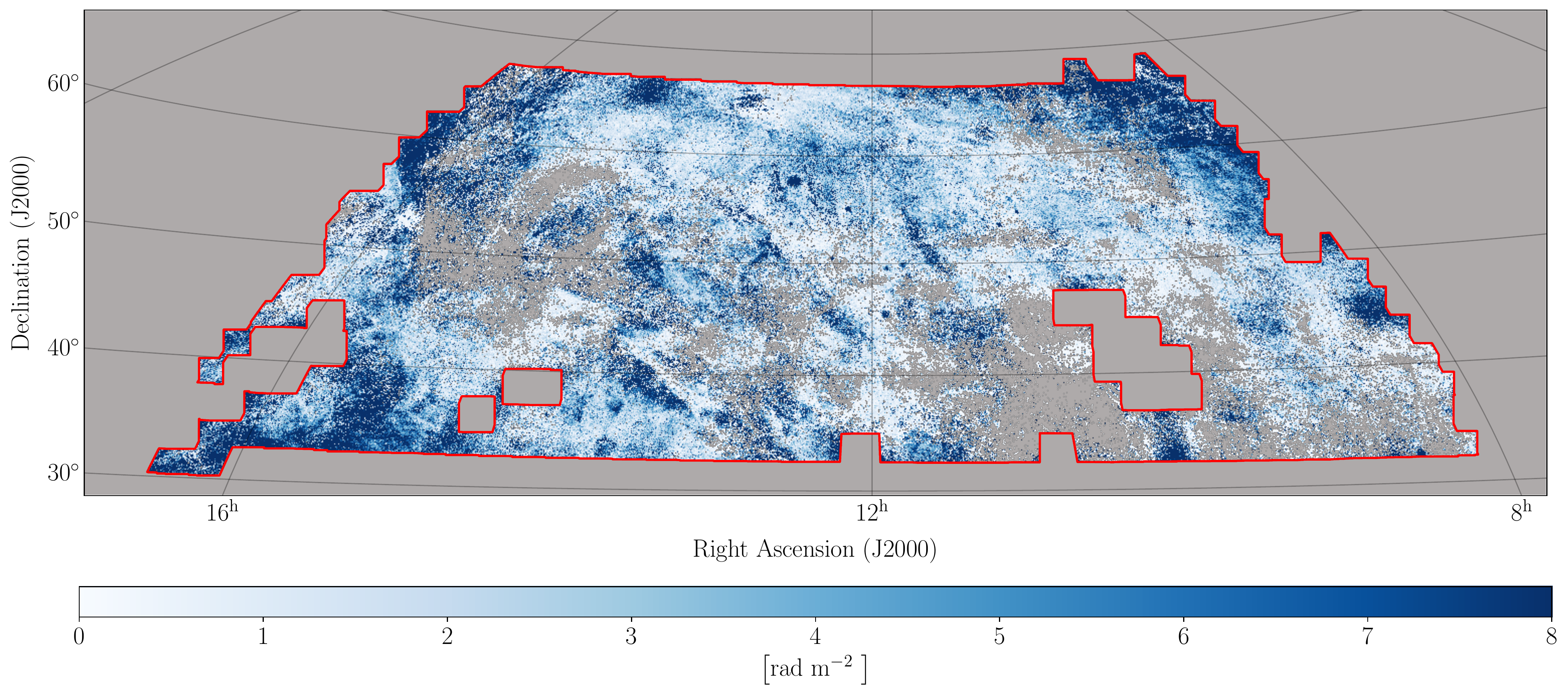}
      \caption{{Moments of the Faraday cubes.} The top panel shows the zeroth moment image representing the total polarised intensity. The middle panel shows the first moment image representing the intensity weighted mean Faraday depth value. The bottom panel shows the square root of the second moment image representing the spread of spectra around first moment values. All images are in equatorial coordinates. The red line outlines mask edges. 
                  }
         \label{Ms}
\end{figure*}

Calculated Faraday moment images ($M_0$, $M_1$, and $M_2$) are presented in equatorial coordinates in Fig.~\ref{Ms}. The zeroth moment shows Loop III as the dominating feature.  We also see small-scale emission in the south-eastern quarter of the mosaic, as mentioned in the previous section. In the first moment image (Fig.~\ref{Ms}, middle panel), the blue colour corresponds to structures at negative Faraday depths, and the red colour corresponds to structures at positive Faraday depths. A large-scale structure at negative Faraday depths in the upper part of the mosaic is connected with Loop III. Consistent, negative Faraday depth implies that the average parallel component of the magnetic field in the loop area is pointing away from us. Below Loop III, the mosaic is dominated by structures at positive Faraday depths. The fact that Loop III remains as striking in zeroth and first moment images as in Fig. \ref{mosaic_colored}, means that it is the dominant feature along these LOSs, contributing the most to the moments.

The second moment image (Fig.~\ref{Ms}, bottom panel) shows that many regions across the mosaic have complex Faraday spectra. However, in the region of Loop III, we do not see a distinct structure. Values in this region are mostly low, indicating simple Faraday spectra with Loop III emission dominating over other emission along the same LOS. The most distinguished structure seen in the image is the zig-zag structure, which was previously described in Section \ref{subsec:3.2}. High values for the second moment along the zig-zag structure indicate complex Faraday spectra, composed of multiple Faraday structures along the Faraday depth. An example of such a spectrum is given in Fig. \ref{chosen_spectra} (spectrum 2), {in which two peaks are clearly visible -- the one closer to zero comes from emission connected to the loop and the other one comes from the zig-zag structure.} Such Faraday spectra can arise from two different physical scenarios. The first is associated with the presence of two or more Faraday thin structures of similar intensities along the LOS. The second is associated with the edges of a Faraday thick structure at these frequencies. To distinguish between the two scenarios, one can use observations at higher frequencies, where structures which are Faraday thick at low radio frequencies, {will become} Faraday thin. In the case of a Faraday thick structure (at low frequencies), we expect the first and the second moments found at the two radio-frequency regimes to be comparable. This comes from the fact that at low frequencies, {we observe} the edges of the Faraday thick structure \citep{brentjens05, vaneck17}. The same structure observed at high frequencies {will become} Faraday thin, and found at a Faraday depth corresponding to the average Faraday depth of the {edges observed} at low frequencies. Another possibility to discriminate between the two aforementioned physical scenarios is to use distance estimations for the structures in Faraday spectra as was done in \citet{vaneck17, thomson19}. 

\section{Comparison with DRAO GMIMS data and the Galactic Faraday Sky map}\label{sec:discussion}

\begin{figure}
   \centering
   \includegraphics[width=\hsize]{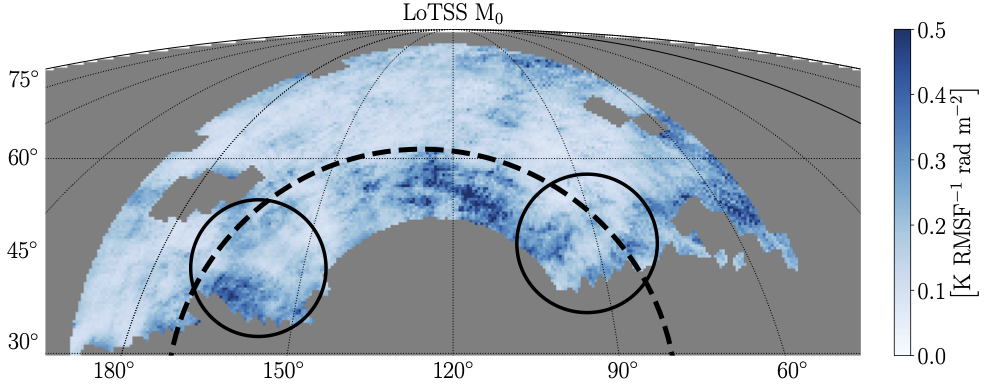}
   \includegraphics[width=\hsize]{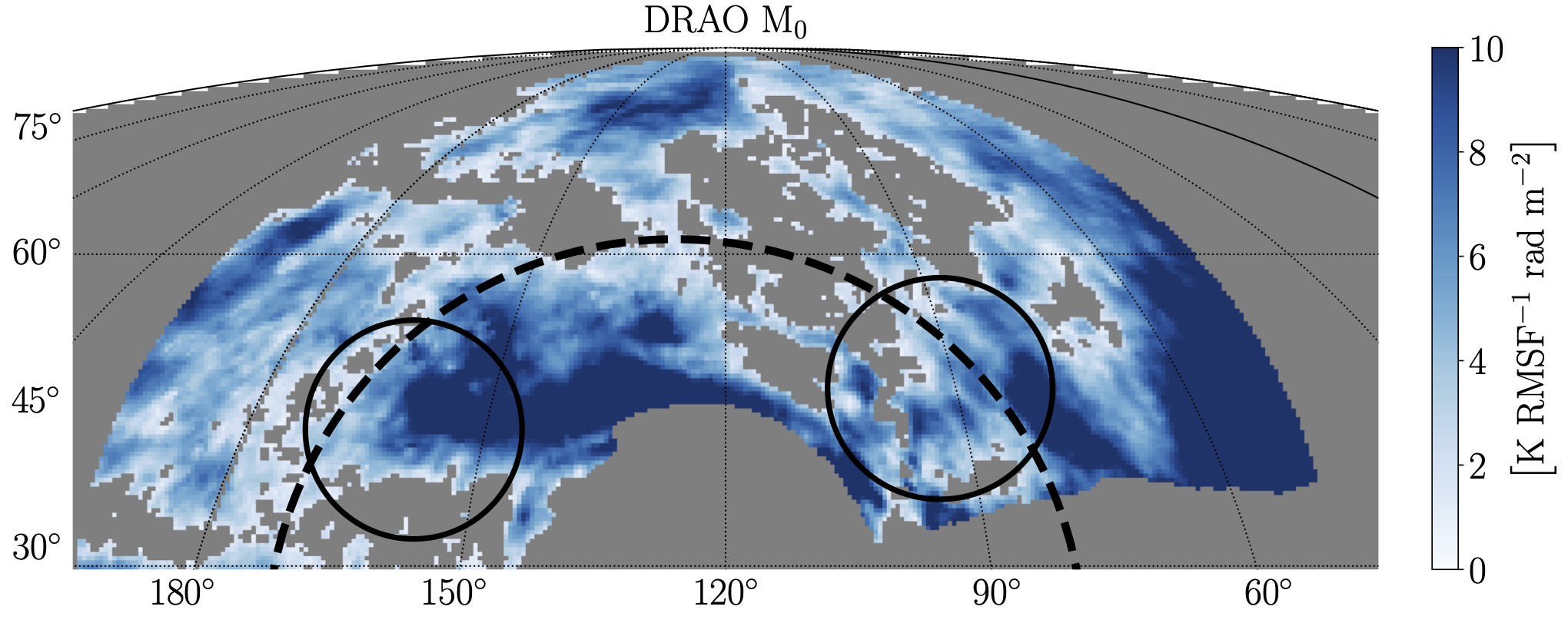}
      \caption{LoTSS and DRAO GMIMS zeroth moment images in Galactic coordinates with the same projection as in Fig. \ref{colored_gal} and they have a resolution of 40 arcmin. Circles mark areas discussed in Section \ref{subsec:5.1}. The units for the LoTSS zeroth moment were converted to $\mathrm{K~RMSF^{-1}~rad~m^{-2}}$ to simplify the comparison with DRAO GMIMS data. {The intensity of 1 $\rm{mJy~PSF^{-1}~RMSF^{-1}~rad~m^{-2}}$ corresponds to $\sim 0.5~\rm{K~RMSF^{-1}~rad~m^{-2}}$.} 
              }
         \label{m0LD}
\end{figure}

 \begin{figure}
   \centering
   \includegraphics[width=\hsize]{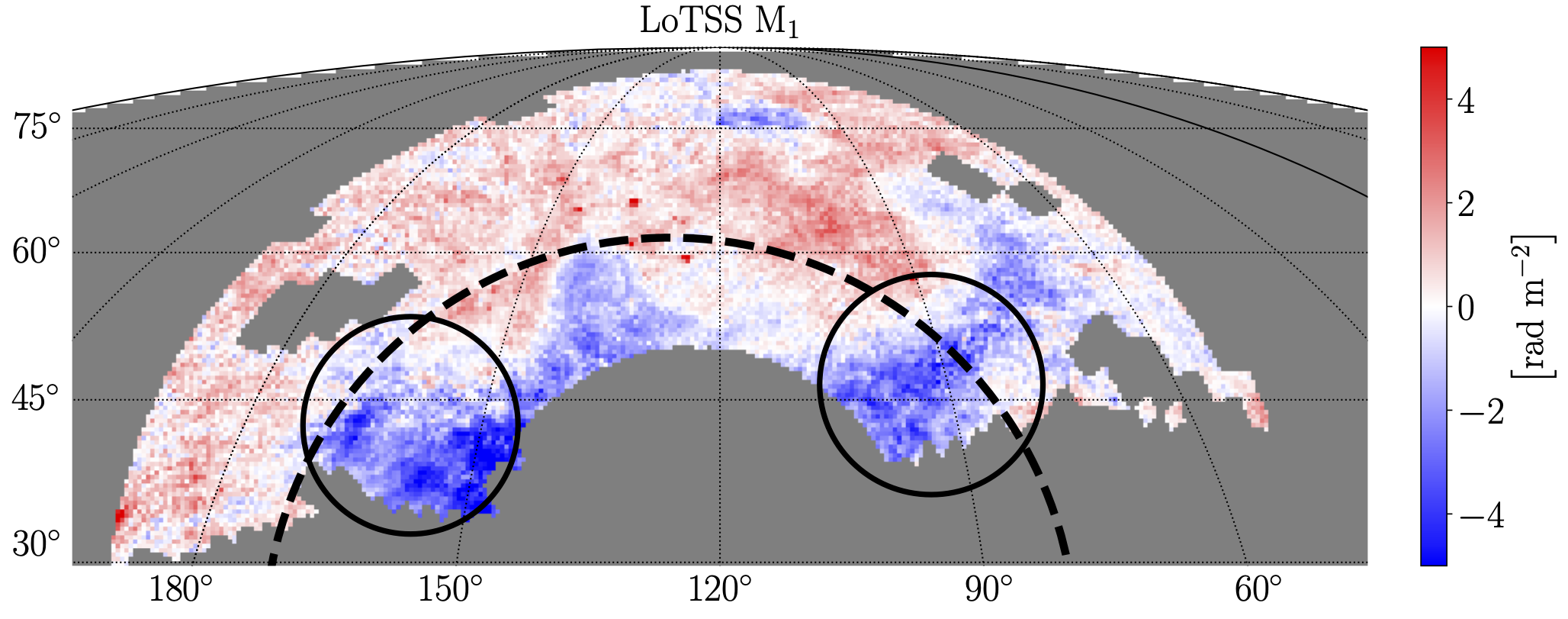}
   \includegraphics[width=\hsize]{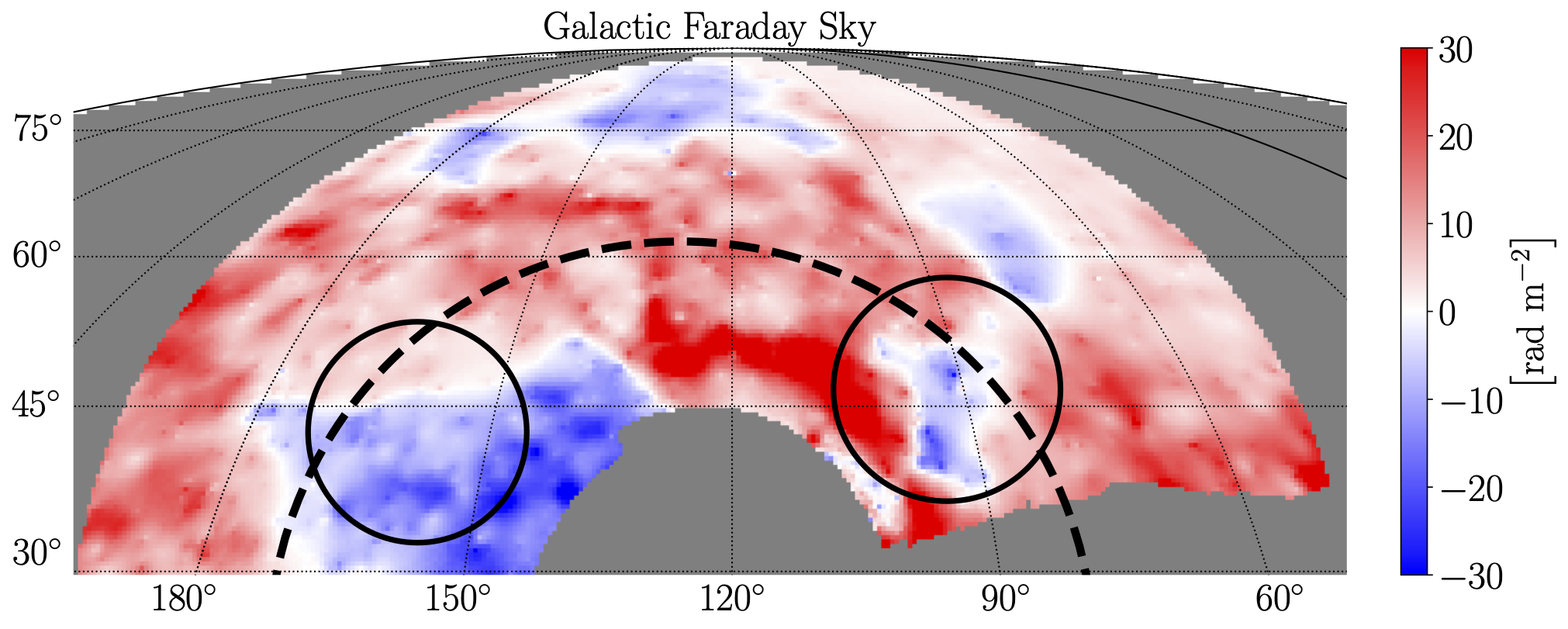} 
   \includegraphics[width=\hsize]{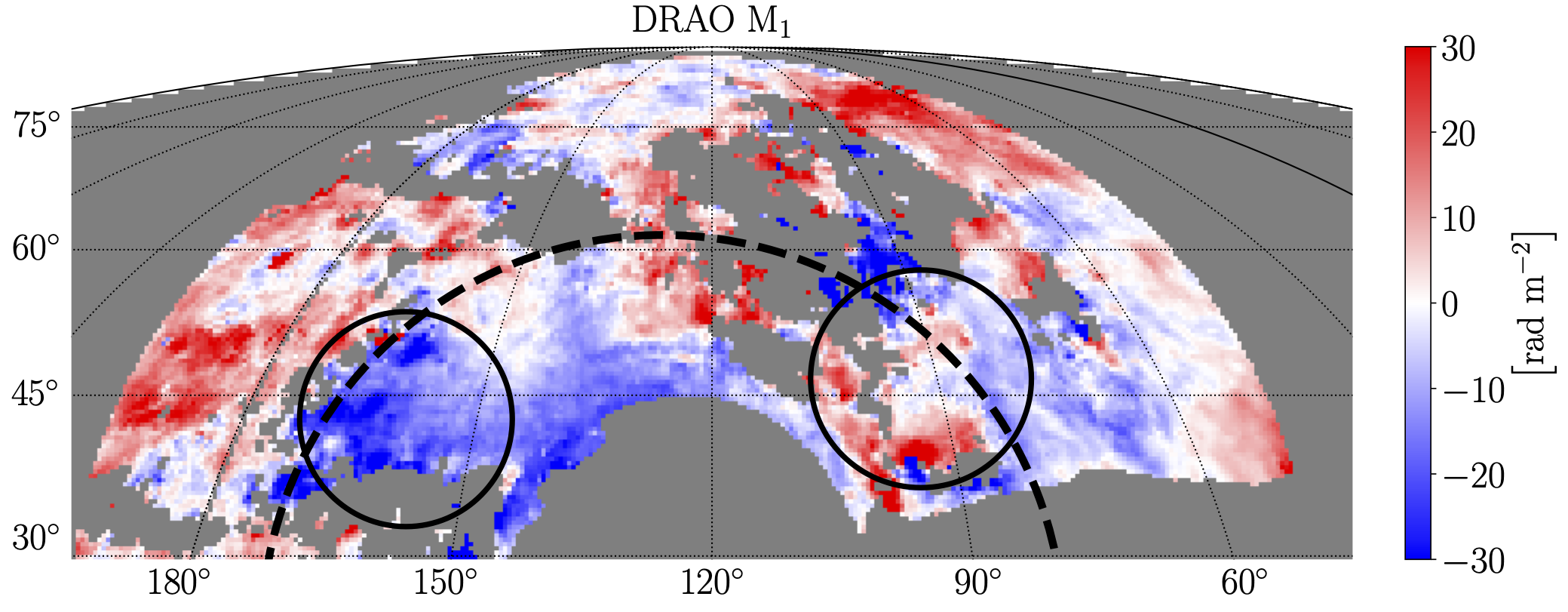}
      \caption{First moment images of LoTSS, {Galactic Faraday Sky, and DRAO GMIMS} are shown in the upper, middle, and lower panels, respectively. The dashed line illustrates the boundary between the lower and upper regions of the shown maps. All images are in Galactic coordinates {with the same projection as in Fig. \ref{colored_gal},} and they have resolution of 40 arcmin. {Circled areas are the same as in Fig. \ref{m0LD}.}
              }
         \label{m1LDG}
\end{figure}

\begin{figure*}
   \centering
   \includegraphics[width=\textwidth]{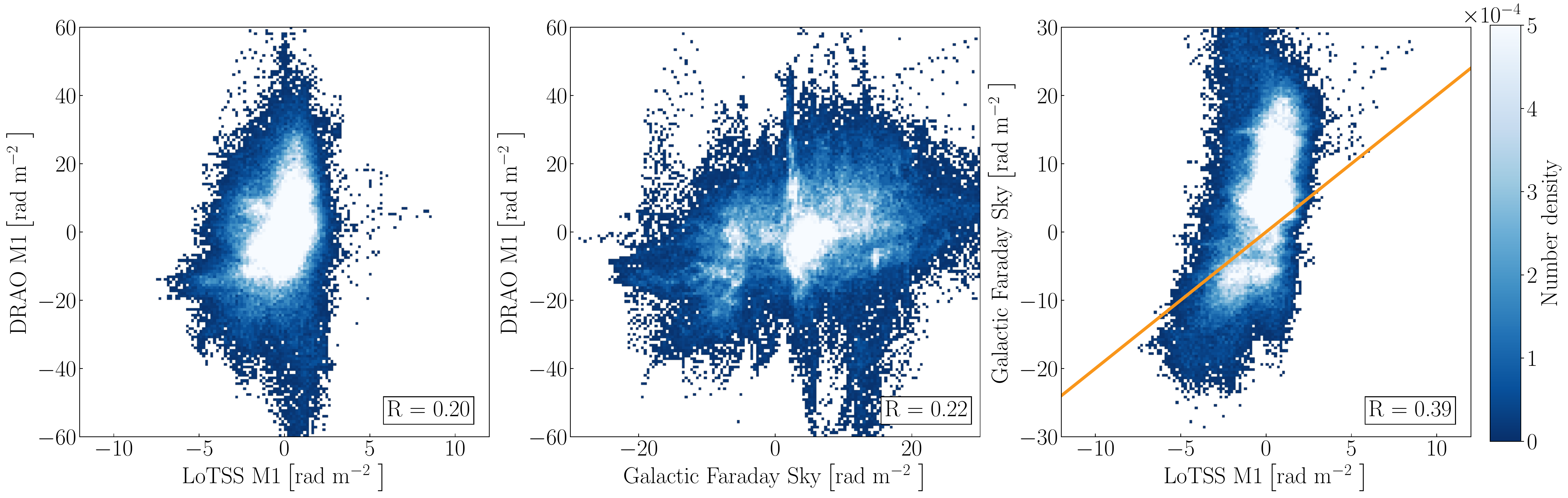}
      \caption{2D histograms of DRAO first moment versus LoTSS first moment (left), {DRAO GMIMS first moment versus Galactic Faraday Sky values} (middle), and Galactic Faraday Sky values versus LoTSS first moment (right). Light blue bins indicate a high number density. The Pearson correlation coefficient, R, is given at the bottom right of each panel. The histograms were made with data at a resolution of 40 arcmin. In the making of all of the histograms, the same area was used.
              }
         \label{2d_hists}
\end{figure*}

\begin{figure}
   \centering
   \includegraphics[width=\hsize]{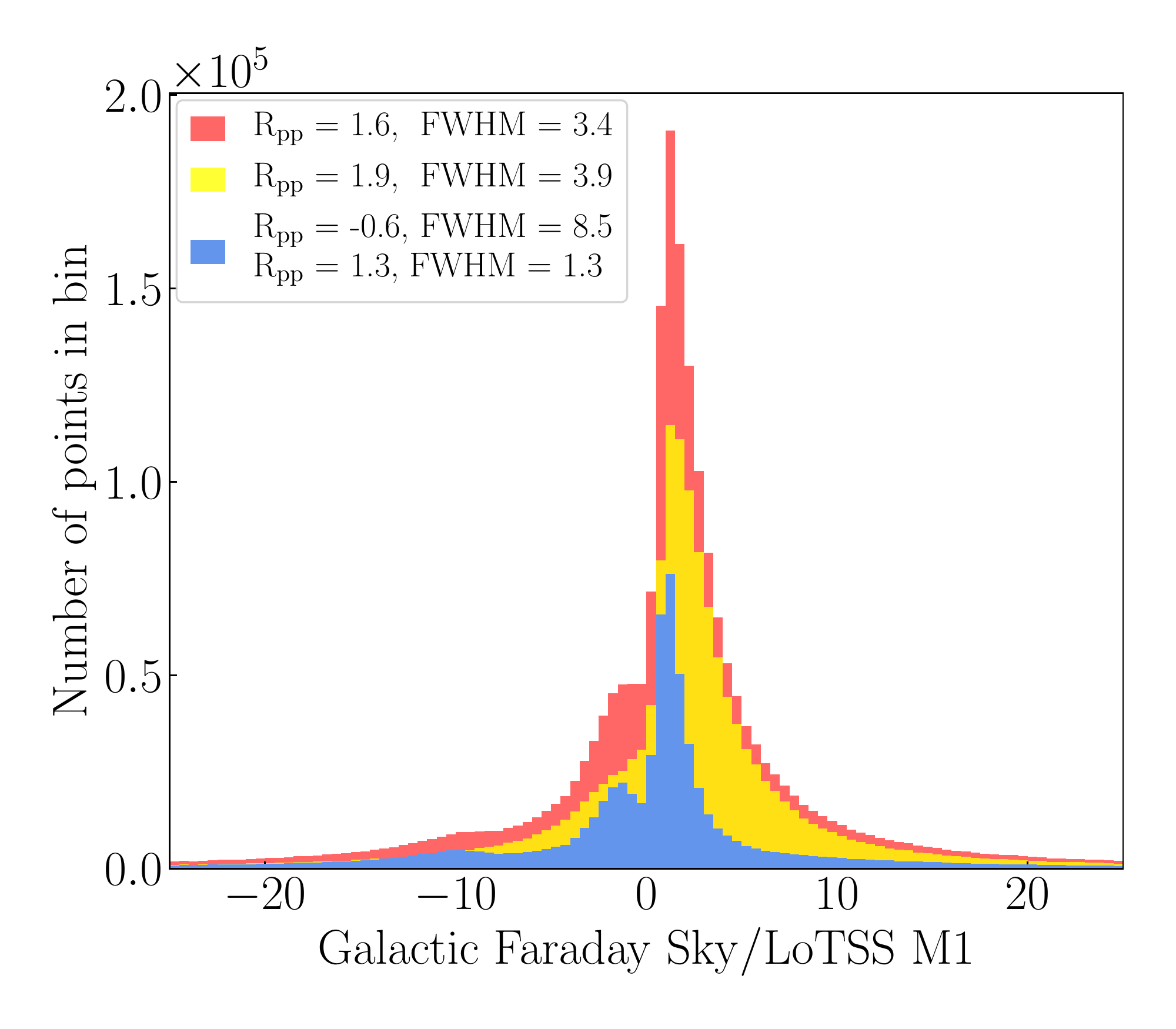}
      \caption{Histogram of the Galactic Faraday Sky to LoTSS first moment ratio in different regions. Blue and yellow histograms display the lower and the upper region, respectively. The red histogram displays both regions combined. We used a Lorentzian fit to characterise the histograms; the obtained ratio peak positions ($\mathrm{R_{pp}}$) and FWHMs are given in the legend above the plot. 
              }
         \label{HLratio_hist}
\end{figure}

\begin{figure*}
   \centering
   \includegraphics[width=\hsize]{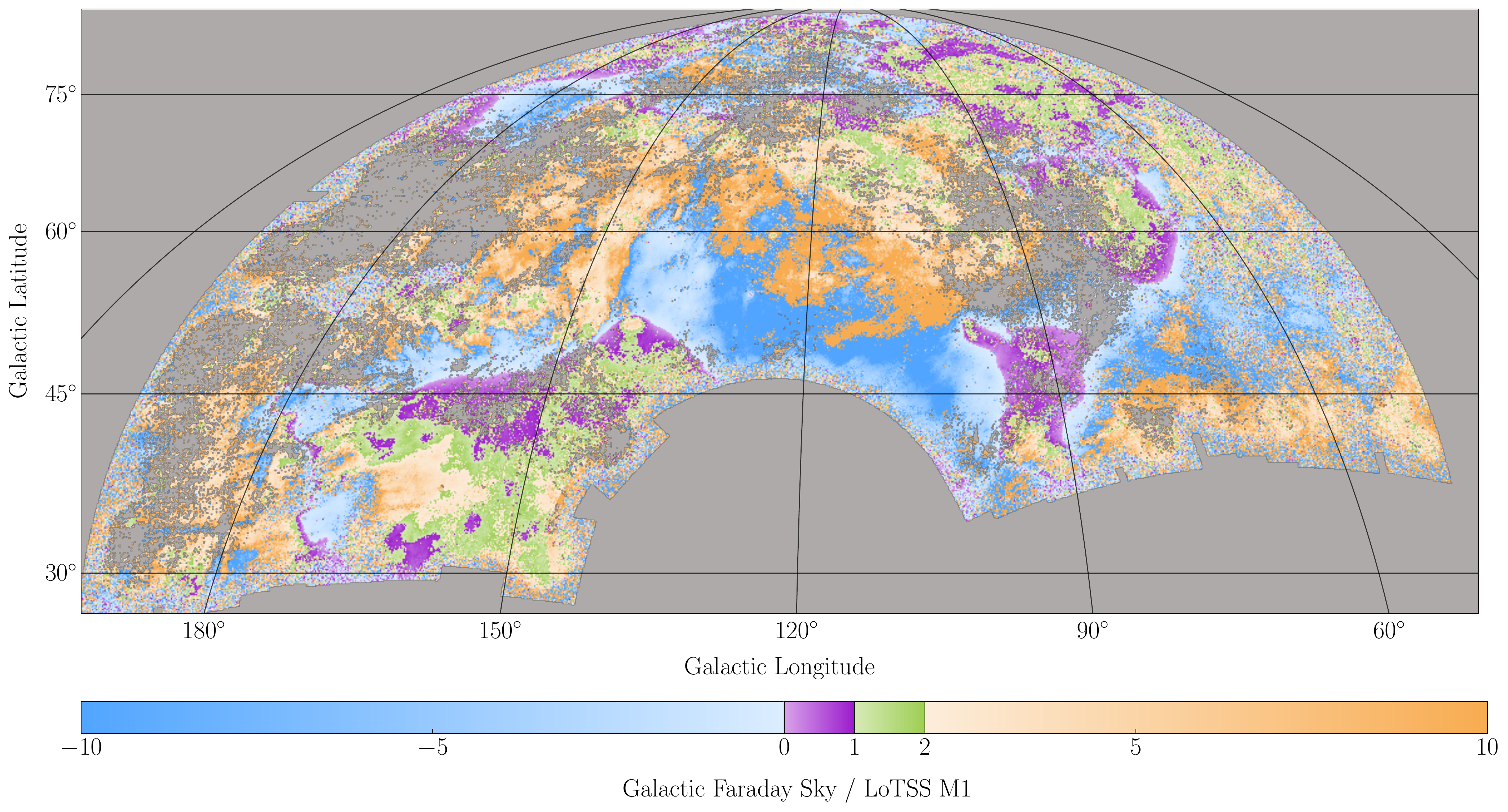}
      \caption{Map of the Galactic Faraday Sky to LoTSS first moment ratio. The blue colour marks the area where the two data sets trace different structures. Purple marks ratios lower than 1, green represents ratios in range from 1 to 2, and orange marks ratios higher than 2.
              }
         \label{ratio_map}
\end{figure*}

We compared our results with the Faraday moment maps presented by \citet{dickey19}, with an angular resolution of 40 arcmin, derived from the preliminary data of DRAO Global Magneto-Ionic Medium Survey (GMIMS) polarisation data set. This polarisation data set is now public through the data release described by \citet{wolleben21}. The Faraday depth resolution of the used DRAO GMIMS moments is $\boldmath{\mathrm{140~rad~m^{-2}}}$ \citep{dickey19}. Structures at the scale of $\mathrm{\sim1~rad~m^{-2}}$ that are seen in LoTSS data {are} resolved out in the DRAO data. On the other hand, some of the Faraday thick structures for LOFAR at these frequencies become Faraday thin and observable. Comparing the two data sets {allows} us to obtain a deeper understanding of the Faraday thickness of the observed structures and the effect of depolarisation on the observed volume of the ISM, both of which are frequency dependent.

Additionally, we compared our results with the Galactic Faraday Sky map \citep{hutschenreuter20, hutschenreuter22}, representing the total RM produced by the Galaxy. As such, this map complements both LOFAR and DRAO GMIMS first moment images in the analysis. The Galactic Faraday Sky was reconstructed using extragalactic sources including, amongst others, the catalogue of polarised sources from the LoTSS survey (O'Sullivan et al. in prep.). We used the publicly available version of the map\footnote{\url{https://wwwmpa.mpa-garching.mpg.de/~ensslin/research/data/faraday2020.html}}, with a resolution of 8.6 arcmin \citep{hutschenreuter22}.

In order to compare different data sets, we smoothed all data to the resolution of {40} arcmin, matching the resolution of DRAO GMIMS moment images. While smoothing, we used the mask from Section \ref{sec:moments}. The smoothing was done in $Q$ and $U$ mosaic cubes, after which the moments with the new resolution were calculated. {To compare other data sets with DRAO GMIMS data,} we excluded additional regions, which were originally masked in DRAO GMIMS moment images (see bottom panel of Fig. \ref{m1LDG}). 

\subsection{Visual comparison}
\label{subsec:5.1}
We performed a visual comparison of the LoTSS and DRAO GMIMS moment images and the Galactic Faraday Sky map. All the images in this subsection are shown in Galactic coordinates to emphasise Loop III.

Visually comparing LoTSS and DRAO GMIMS zeroth moment images, we {find} some indications of emission in DRAO GMIMS data that could be connected to Loop III in the areas marked with black circles in Fig.~\ref{m0LD}. In the circled area on the left of the DRAO GMIMS image (Fig.~\ref{m0LD}{, bottom panel}), we see structures that seem to follow the shape of the loop, while the same area in the LoTSS zeroth moment image has an intensity two times lower than the loop average intensity. Both DRAO GMIMS and LoTSS first moment values in this area are negative and have comparable values {(Fig. \ref{m1LDG})}. {This result suggests that Loop III is also visible in the DRAO GMIMS data. The reason for the differences in what we see in the two frequency regimes may lie either in different Faraday thicknesses of observed structures or frequency-dependent depolarisation effects. If the structure we are seeing is Faraday thin in both regimes, it could have a lower intensity at low frequencies due to a stronger effect of depolarisation. We discuss this in more detail in the following subsection.}

In the {circled area on the right} (Fig.~\ref{m0LD}), we observe {some} emission in the DRAO GMIMS zeroth moment. {This} area again shows lower-than-average intensity in the LoTSS zeroth moment image; {however,} the morphology does not coincide with the morphology seen in the DRAO GMIMS image. Furthermore, we do not see any visual correspondence between the first moment images of DRAO GMIMS and LoTSS in the area of the circle on the right shown in Fig. \ref{m1LDG}. DRAO GMIMS values are {mostly} positive, LoTSS values are {mostly} negative, and the morphology of detected structures is different. We conclude that the loop is not detected in this part of the DRAO GMIMS image. 

There are several possibilities that could explain why the structure detected in DRAO GMIMS is not seen in LoTSS. The structure detected in the DRAO GMIMS images could be further away and depolarised in the LoTSS data. Alternatively, it could be a structure with low brightness and one that is Faraday thick for LoTSS, which would contribute to the LoTSS first moment values negligibly because of the intensity weighting.

As shown in the middle panel of Fig.~\ref{m1LDG}, the map of the Galactic Faraday Sky has negative values of RM in the circled area on the left. The feature has a shape similar to the structure of Loop III seen in the LoTSS first moment image (upper panel of Fig.~\ref{m1LDG}). However, {the circle on the right contains an} interesting shape {stretching over} the north-western area of the loop. {This shape is roughly centred at Galactic coordinates $l=117^\circ$, $b=50^\circ$ and} it has high positive RM values, with a mean around +25 $\mathrm{rad \ m^{-2}}$. The same structure was also present in the first Faraday Sky map by \citet{oppermann12}, where it was associated with an intermediate-velocity arch of HI spectroscopic data. This area in the DRAO GMIMS first moment image has a very low S/N due to depolarisation. In LoTSS, this structure is also not present, which is probably also due to depolarisation.

\subsection{Statistical comparison}
To make a quantitative comparison between different data sets, we represent the data in the form of 2D histograms and calculate the Pearson correlation coefficient. In Fig.~\ref{2d_hists} we show a 2D histogram of DRAO GMIMS versus LoTSS first moment images in {the} left panel, DRAO GMIMS versus Galactic Faraday Sky map in {the} middle panel, and the Galactic Faraday Sky map versus LoTSS in {the} right panel. The strongest correlation is found between LoTSS and the Galactic Faraday Sky, with a correlation coefficient of $R=0.39$. In the other two cases, we did not find a significant correlation between the compared data sets.

The lack of correlation between DRAO GMIMS and LoTSS first moment images ($R=0.20$) can be understood as a result of frequency-dependent Faraday depth resolution and depolarisation. Faraday depth resolution at LOFAR frequencies is two orders of magnitude higher than at DRAO GMIMS frequencies. This makes low-frequency data sensitive to many Faraday thin structures that are not resolved by {the} RMSF function  at higher frequencies due to poor resolution. In addition, depolarisation associated with Faraday rotation is stronger at low frequencies than at high frequencies. Therefore, we are most likely not probing the same volume of the Galaxy in the LoTSS and DRAO GMIMS surveys. Differences between the two data sets may also arise from the different angular scales to which the two telescopes are sensitive. DRAO GMIMS data have poorer angular resolution, which can result in stronger beam depolarisation. On the other hand, DRAO GMIMS data were observed using a single dish telescope that can detect larger scales to which LOFAR is not sensitive.

In the case of DRAO GMIMS and Galactic Faraday Sky, one would expect to see a correlation, as indicated by the results of \citet{dickey19} and \citet{ordog19}. As mentioned before, depolarisation is not expected to be significant at high frequencies. The volume probed should be more similar to the total volume probed by extragalactic sources. However, in this specific region of the sky, the dominating structure (Loop III) is considered to be fairly nearby and it is only partially visible in the DRAO GMIMS data. This is likely due to confusion with other Faraday structures present in the spectra and poor Faraday resolution at higher frequencies and/or beam depolarisation. This could explain the lack of correlation in this specific region ($R=0.22$).

{We note that when smoothing to 40 arcmin, we lost a considerable amount of LoTSS data. To test if the correlation between LoTSS M1 and the Galactic Faraday Sky persists at a LoTSS native resolution, we oversampled the Galactic Faraday Sky and repeated the analysis. We obtained a Pearson correlation coefficient of 0.43, which shows that smoothing the data does not change the correlation significantly. Nevertheless, we continued to investigate this correlation at a LoTSS native resolution.}

To quantify {this correlation}, we plotted a histogram of the ratio between the Galactic Faraday Sky and the LoTSS first moment (see Fig. \ref{HLratio_hist}). The peak of the distribution, $R_{pp}$, represents the average ratio found in the analysed part of the sky and is found to be at {1.6}, while the FWHM represents the spread of the distribution, and it is {3.4}. Although the spread is rather large, this value is similar to the value of  2, expected for the 'Burn slab model' \citep{burn66}. This model assumes a uniform magnetic field and electron density throughout the slab, where each layer of the slab is emitting and Faraday rotating synchrotron emission. As a result of depth depolarisation, the expected Faraday depth values are half that of the total RM through the slab. The observed discrepancy suggests more complex physical conditions than assumed in this simple model.

We separated the mosaic in two parts {to examine the ratios in two distinct regions}. The lower region mostly contains the polarised emission associated to Loop III and the upper region contains the polarised emission at higher Galactic latitudes, above the loop. The boundary between the two regions is illustrated with a dashed line in Fig.~\ref{m1LDG}. The histograms of the Galactic Faraday Sky to the LoTSS first moment ratio for the lower and upper region are given in Fig. \ref{HLratio_hist}. The ratio distribution in the lower region shows two peaks, a higher one with a positive value and a lower one with a negative value corresponding to areas where the compared data have opposite signs. The positive and the negative {peaks can be described by two Lorentzians centred at 1.3 and -0.6,} respectively, with FWHMs of {1.3 and 8.5}, respectively. The negative peak comes from the areas that do not correlate -- the Faraday depths of structures seen there have opposite signs. In the upper region, we found {a peak at 1.9, with a FWHM of 3.9.}

\subsection{Insights from simple toy models}

To understand the ratios in different regions, we ran synthetic observations of LOFAR data with a few simple toy models (variations of the Burn model) of different physical scenarios. In the models, we assumed that the Faraday rotation is present along the full LOS and that the magnetic field and electron density are uniform. Models are distinguished by the location and thickness of the emitting regions (for details see Appendix \ref{app:burn}.). Similar toy models were considered by \citet{ordog19}. An important difference between our simple toy models and LoTSS data is that the models represent a single LOS, while LoTSS data come from multiple lines of sight that could be probing very different ISM configurations which correspond to different models (i.e. the double peaked distribution of Fig. \ref{HLratio_hist}). 

The first toy model (Model A) is a Burn slab with emission and rotation along the full LOS, {which is the same as that in \citet{burn66}}. In this limit, we obtained a ratio of 2 {between the modelled total Galactic RM, as it would be probed by extragalactic sources, and the observed polarised emission}. In the second model (Model B), we distributed one or more synchrotron emitting layers along the LOS with the Faraday rotating medium in between, acting as Faraday screens -- in this case, there is no differential Faraday rotation. We further expanded this model into three cases regarding the location of the emitting layers. In Model B.0, we considered a single emitting layer followed by a rotating volume acting as a single screen. The first moment value is then the same as the total RM value, and the ratio is equal to 1. In Model B.1, multiple emitting layers are distributed along the full LOS and we got ratios between 1 and 2. In Model B.2, the emitting layers are distributed in the closer half of the LOS, and the ratios obtained are higher than 2. In the third model (Model C), we attenuated the emission in the Burn slab with increasing distance from the observer. In this case, the ratio obtained exceeds the value of 2 and grows higher with a decreasing amount of the LOS being affected by differential Faraday rotation.

These toy models {help} us to develop an intuition on the variation of the Galactic  Faraday  Sky  to  LoTSS  first  moment ratios seen in the data. {The positive ratios obtained for the full mosaic area and its lower and upper parts could correspond} to model B.1. However, in the context of a large distribution spread, the results could correspond to any of the toy models presented here, or a combination of them. 

Since averaging over large areas of the data, with more than one distinct feature, may be too crude, we present a map of the Galactic Faraday Sky to LoTSS $M_1$ ratios in Fig. \ref{ratio_map}. From this map, we can understand the presence of a negative peak found in the lower region of the mosaic as a signature of the intermediate velocity cloud seen at high positive latitudes only in the Galactic Faraday Sky map. Negative ratios (blue in Fig. \ref{ratio_map}) can also be attributed to a flip in the magnetic field direction along the LOS -- the local magnetic field can have a sign opposite from the average along the full Galaxy. The flip in magnetic field direction can also produce ratios from 0 to 1 (purple in Fig. \ref{ratio_map}), magnetic fields along the full LOS can average out {to low values}, while the nearby magnetic field is stronger. Ratios from 1 to 2 are marked in green and indicate our model B.1. Orange areas of the map show positive ratios, higher than 2. Along the LOS where the ratios are much higher than 2, there is much more depolarisation than is predicted by the Burn slab model.

Our toy models suggest that these ratios could carry information about the relative distribution of synchrotron emitting and Faraday rotating regions along the sight line. Actual observations, however, may be the composite result of multiple scenarios at play which reflect the complexity of the Faraday sky. Further analysis, which is beyond the scope of this first paper, will be done in order to reach an in-depth understanding of the different regions under study.

\section{Summary and conclusions}\label{sec:end}

In this paper, we have produced a mosaic Faraday cube combining 440 observations from the LOFAR Two-metre Sky Survey DR2. The mosaic covers around 3100 square degrees in the high-latitude outer Galaxy, which makes it the biggest low-frequency polarisation mosaic in the northern sky to date. The mosaic has an angular resolution of 4.3 arcmin and a Faraday depth resolution of $\sim 1~\mathrm{rad~m^{-2}}$.

The mosaic revealed the richness of diffuse polarised emission morphology. Depolarisation canals of different lengths, widths, and orientations fill the area of the mosaic. Lower Galactic latitudes {are} dominated by large angular scale (tens of degrees) polarised emission associated with Loop III, while higher Galactic latitudes show patchy and more diffuse polarised emission, organised in features on smaller angular scales of 1 degree.

The big gradient {in Faraday depth} spreads over $30~\mathrm{rad~m^{-2}}$ in the direction outward from the centre of Loop III, covering tens of degrees. This gradient is probably associated with a shock front that could have been formed by one or more supernova explosions in the Galactic disk \citep{berkhuijsen1971a, kun07}. 

We have further analysed polarised emission by using zeroth, first, and second Faraday moment images and we compared them to {moments produced from the DRAO GMIMS data set} at 1.4 GHz \citep{dickey19} and to the Galactic Faraday Sky map \citep{hutschenreuter22}. We see some morphological similarities between zeroth moment DRAO GMIMS and LoTSS images in the region of Loop III. In other areas, we see no such correlation, which can be explained by a different Faraday depth resolution and a different maximum Faraday scale to which each survey is sensitive. 

We have found a correlation between the LoTSS first Faraday moment image and the Galactic Faraday Sky. We have studied this correlation by inspecting histograms of the ratio of the two data sets and by testing several toy models in an effort to explain the obtained ratios. The area dominated by the loop shows high complexity and cannot be properly described by the simple models we present. Areas not dominated by Loop III emission indicate depolarisation of emission coming from great distances. 

From the map of ratios (Fig. \ref{ratio_map}), we can conclude that in the low frequency data, not many lines of sight can be described by the Burn slab model, although this does not seem to apply to higher frequency data \citep{ordog19}. A high sensitivity and strong effect of Faraday rotation at low frequencies result in highly complex Faraday structures along different LOSs. Additionally, this map reveals areas that have negative ratios and do not correlate, meaning that {the compared data sets are probing different structures} (e.g. the aforementioned region in the Galactic Faraday Sky map associated to an intermediate-velocity arch of atomic hydrogen gas). Alternatively, since the two data sets do not probe the same volume, negative ratios and positive ratios lower than 1 can be caused by a magnetic field reversal along the LOS in these specific areas.

This paper only scratches the surface of understanding the observed diffuse polarised emission morphology in the LoTSS data and the complexity of the Faraday spectra. To understand the underlying processes in the ISM, multi-tracer analysis is needed. While such analyses of Faraday tomographic data at low-radio frequencies in smaller areas of the sky have already found connections between different phases of the ISM and the magnetic field \citep{zaroubi15, lenc16, vaneck17, jelic18, bracco20, turic21}, the next step to be taken is testing these findings over larger sky areas. A multi-frequency analysis of the LoTSS data set will be presented in a follow-up paper.

\begin{acknowledgements}
We thank the anonymous referee for constructive comments that improved this manuscript. We thank Mike W. Peel for providing the updated version of the map of polarised synchrotron emission
from WMAP and Planck data before publication. AE, VJ and LT acknowledge support by the Croatian Science Foundation for a project IP-2018-01-2889 (LowFreqCRO) and additionally VJ and LT for the project DOK-2018-09-9169. MH acknowledges funding from the European Research Council (ERC) under the European Union's Horizon 2020 research and innovation programme (grant agreement No 772663). AB acknowledges the support from the European Union’s Horizon 2020 research and innovation program under the Marie Skłodowska-Curie Grant agreement No. 843008 (MUSICA). This paper is based on data obtained with the International LOFAR Telescope (ILT) under project codes $\rm{LC2\_038}$ and $\rm{LC3\_008}$. LOFAR (van Haarlem et al. 2013) is the Low Frequency Array designed and constructed by ASTRON. It has observing, data processing, and data storage facilities in several countries, that are owned by various parties (each with their own funding sources), and that are collectively operated by the ILT foundation under a joint scientific policy. The ILT resources have benefitted from the following recent major funding sources: CNRS-INSU, Observatoire de Paris and Université d'Orléans, France; BMBF, MIWF-NRW, MPG, Germany; Science Foundation Ireland (SFI), Department of Business, Enterprise and Innovation (DBEI), Ireland; NWO, The Netherlands; The Science and Technology Facilities Council, UK; Ministry of Science and Higher Education, Poland. This research made use of Montage. It is funded by the National Science Foundation under Grant Number ACI-1440620, and was previously funded by the National Aeronautics and Space Administration's Earth Science Technology Office, Computation Technologies Project, under Cooperative Agreement Number NCC5-626 between NASA and the California Institute of Technology. MJH acknowledges support from the UK Science and Technology Facilities Council [ST/V000624/1]. AD acknowledges support by the BMBF Verbundforschung under the grant 05A20STA. The J{\"u}lich LOFAR Long Term Archive and the German LOFAR network are both coordinated and operated by the J{\"u}lich Supercomputing Centre (JSC), and computing resources on the supercomputer JUWELS at JSC were provided by the Gauss Centre for supercomputing e.V. (grant CHTB00) through the John von Neumann Institute for Computing (NIC). 

\end{acknowledgements}

\bibliographystyle{aa}
\bibliography{reference_list} 

\begin{appendix}

\onecolumn{
\section{Zoom-ins}\label{app:zoom}
In this section we present the selected zoom-ins on the mosaic in Galactic coordinates.

\begin{figure}[h]
   \centering
   \includegraphics[width=0.95\textwidth]{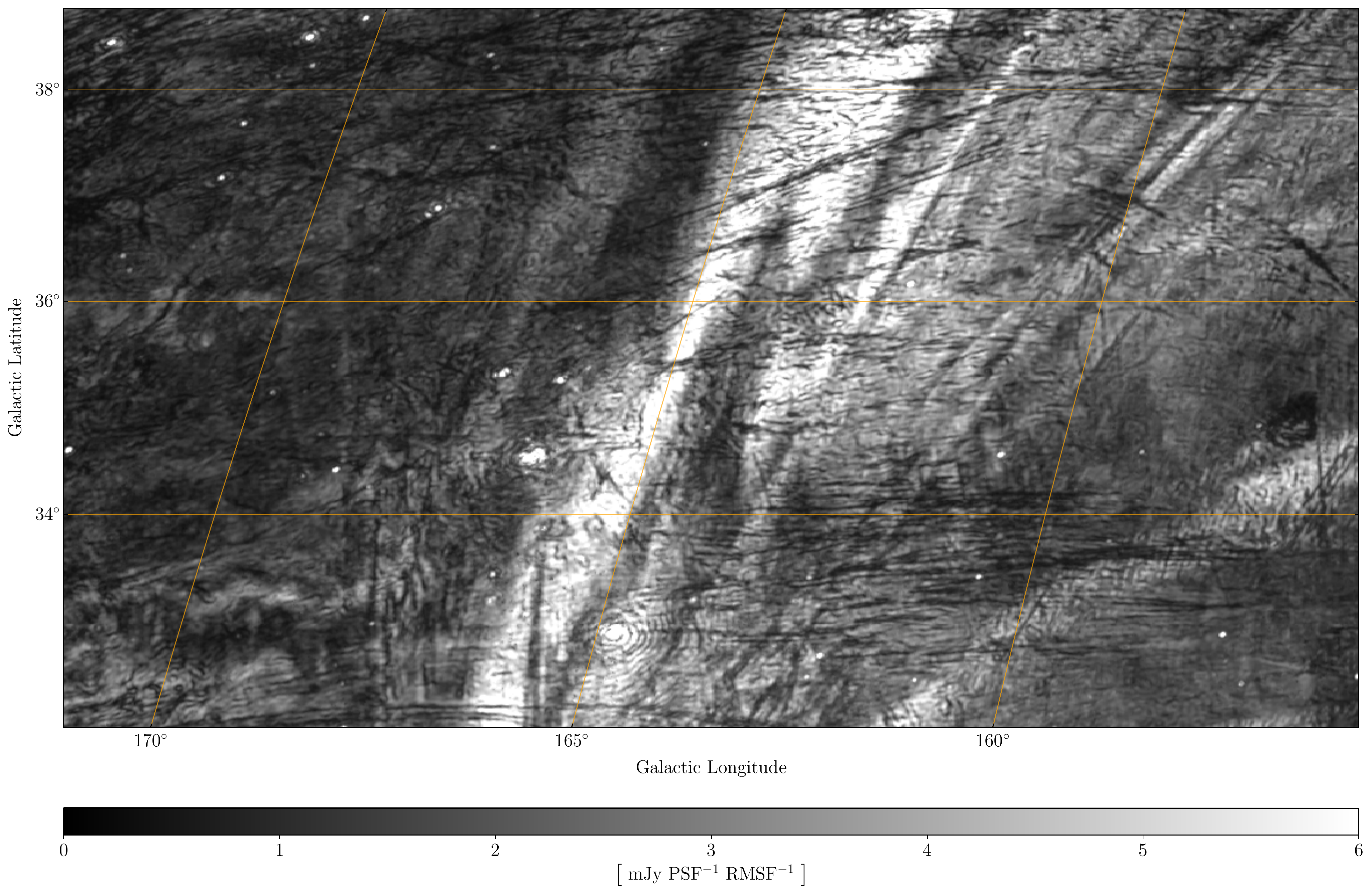}
      \caption{Zoom 1. Magnified maximum polarised intensity image, covering $3\%$ of total mosaic area. This image zooms in on depolarisation canals following Loop III, as well as the ones perpendicular to the loop. 
              }
         \label{zoom1}
\end{figure}

\begin{figure}[h]
   \centering
   \includegraphics[width=0.95\textwidth]{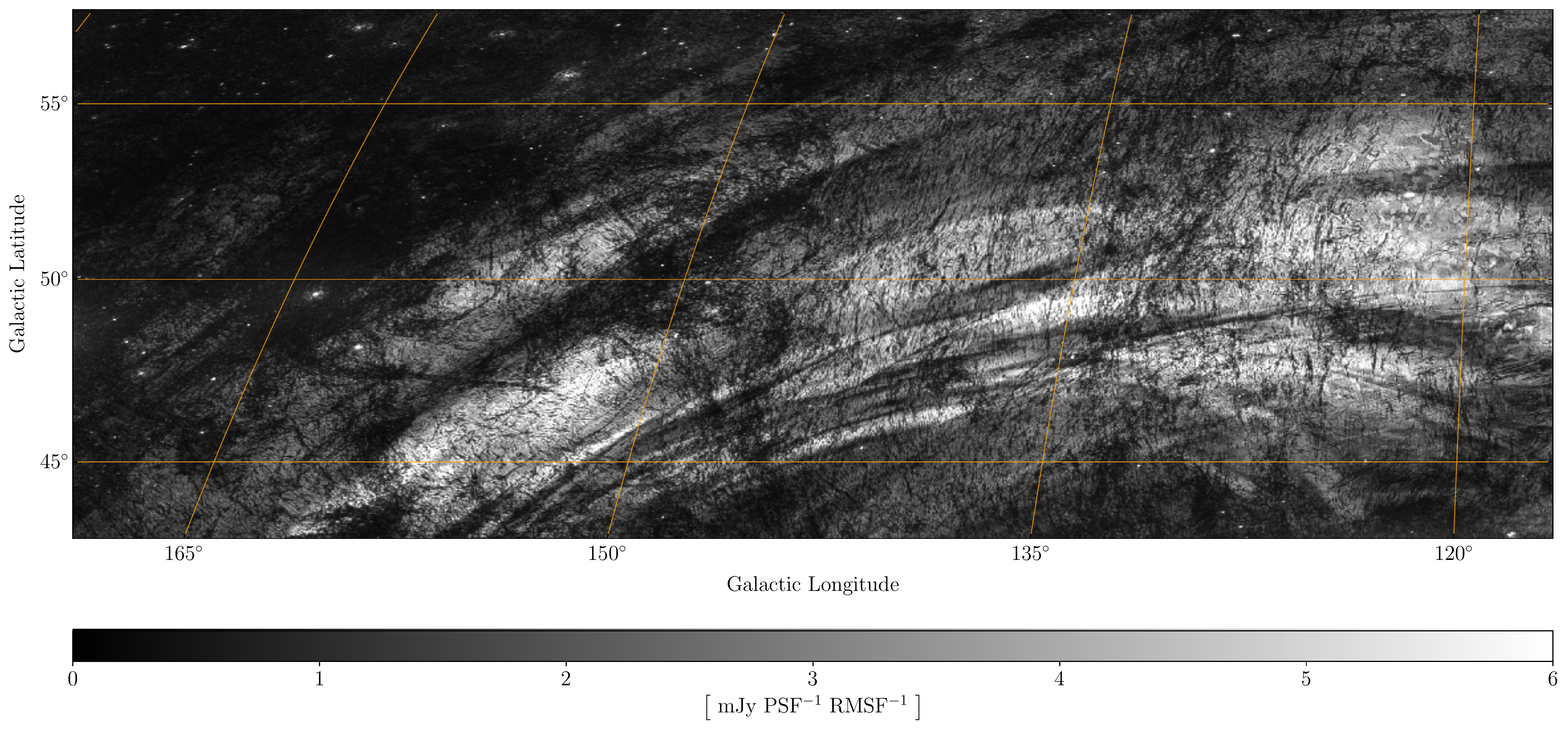}
      \caption{Zoom 2. Magnified maximum polarised intensity image, covering 17\% of total mosaic area. This image shows depolarisation canals stretching along Loop III.
              }
         \label{zoom2}
\end{figure}

\begin{figure}[h]
   \centering
   \includegraphics[width=\textwidth]{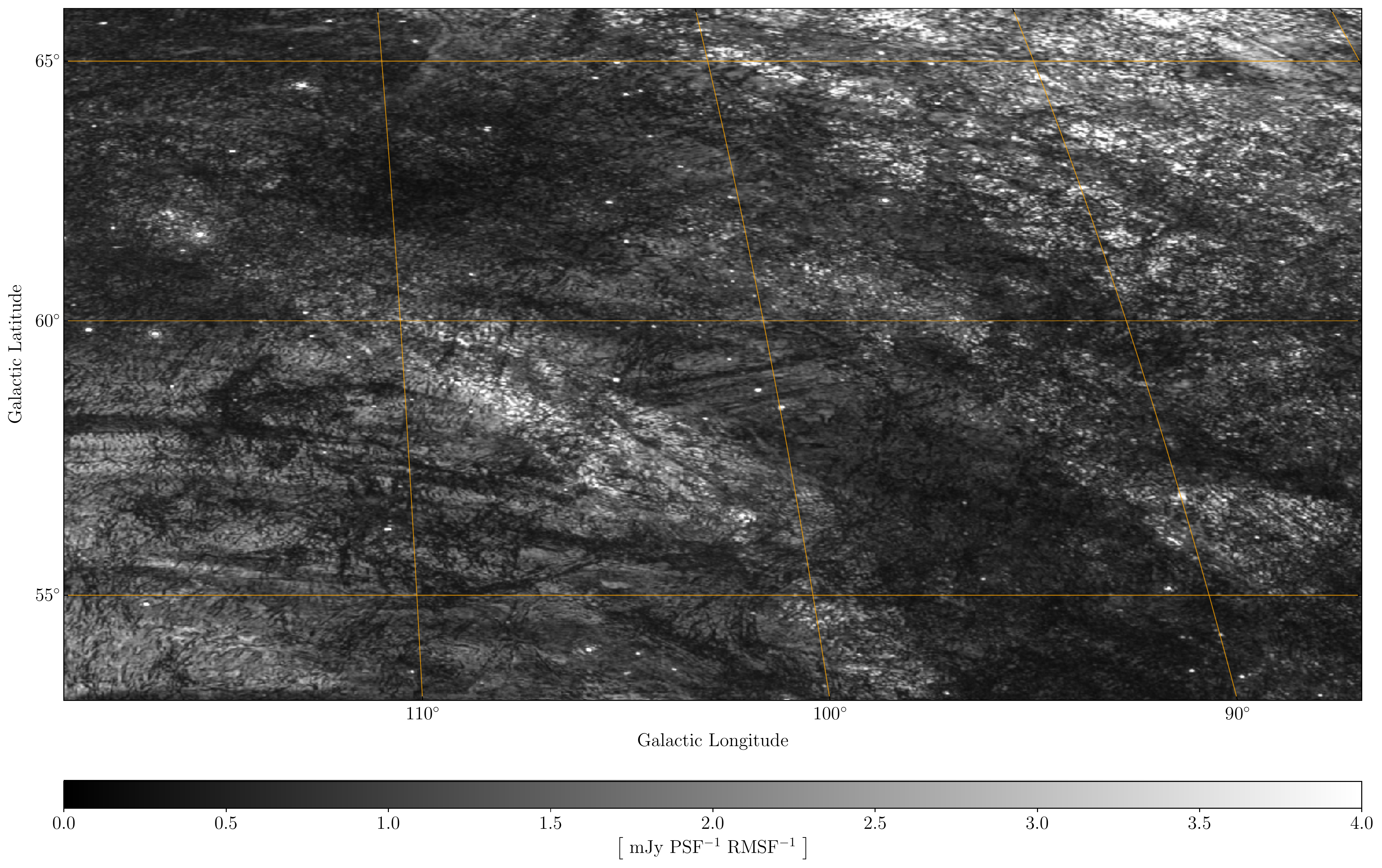}
      \caption{Zoom 3. Magnified maximum polarised intensity image, covering 7\% of total mosaic area.
      The image shows the emission coming from the zig-zag structure mixed with Loop III emission (coming from the bottom left corner).
              }
         \label{zoom3}
\end{figure}
}

\onecolumn{
\section{Mosaic at chosen Faraday depths}\label{app:mosaic_slices}

In this appendix, we present the selected slices from the mosaic Faraday cube. We show only the images at Faraday depths from -25 $\mathrm{rad~m^{-2}}$ to +15 $\mathrm{rad~m^{-2}}$, since there is almost no emission outside of this range. Figures \ref{-25_-15} and \ref{-10_0} mainly show the Loop III emission. The zig-zag structure discussed in Sect. \ref{subsec:3.2} can best be seen in images at 5 and 10 $\mathrm{rad~m^{-2}}$ in Fig. \ref{0_15}.
\begin{figure}[!h]
   \centering
   \includegraphics[width=0.89\textwidth]{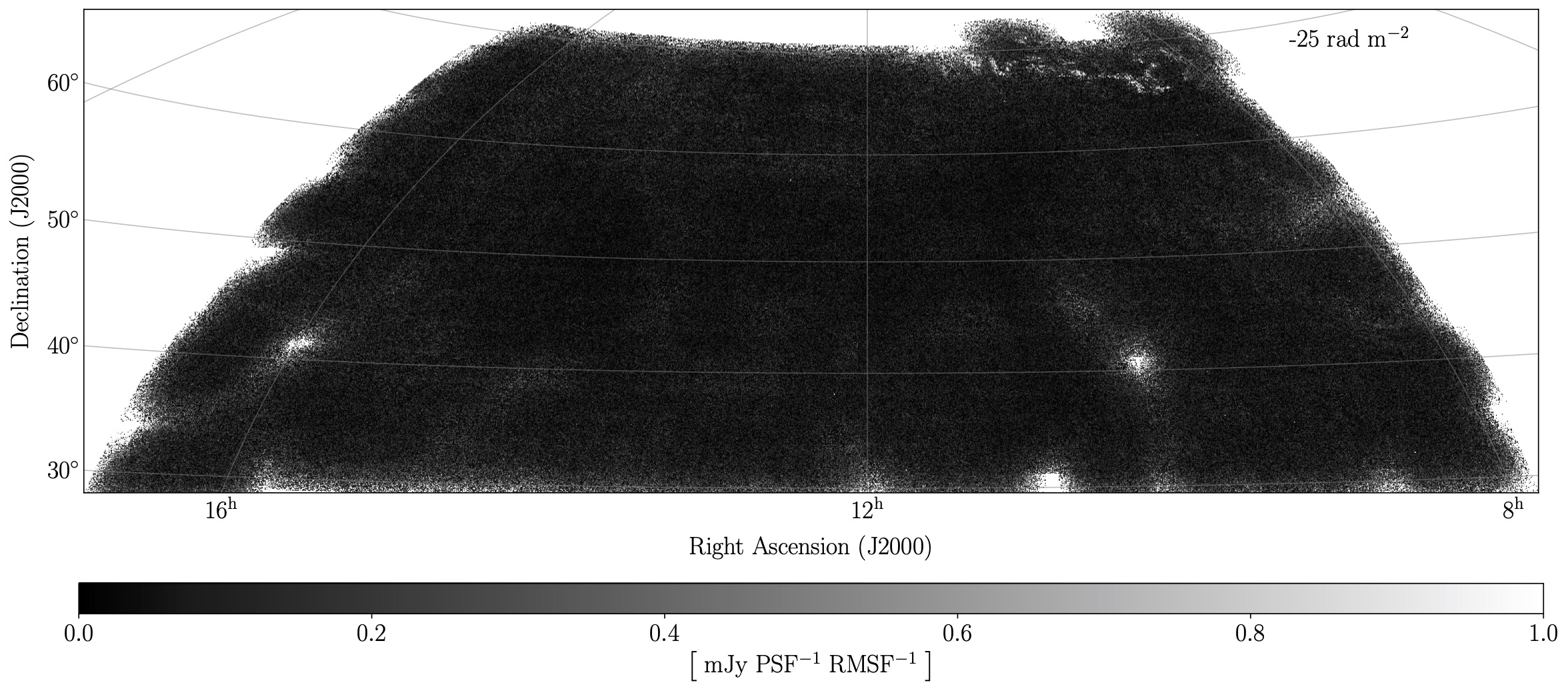}

   \includegraphics[width=0.89\textwidth]{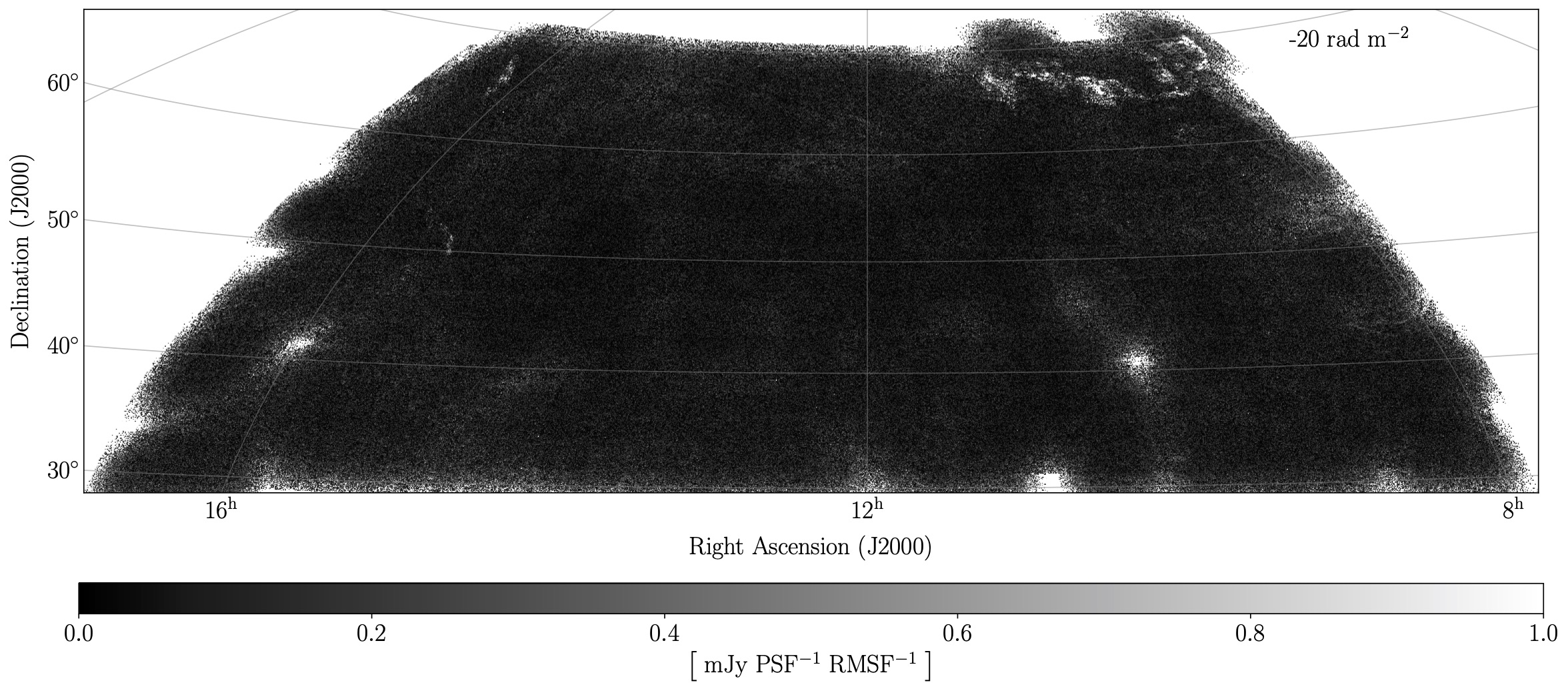}

   \includegraphics[width=0.89\textwidth]{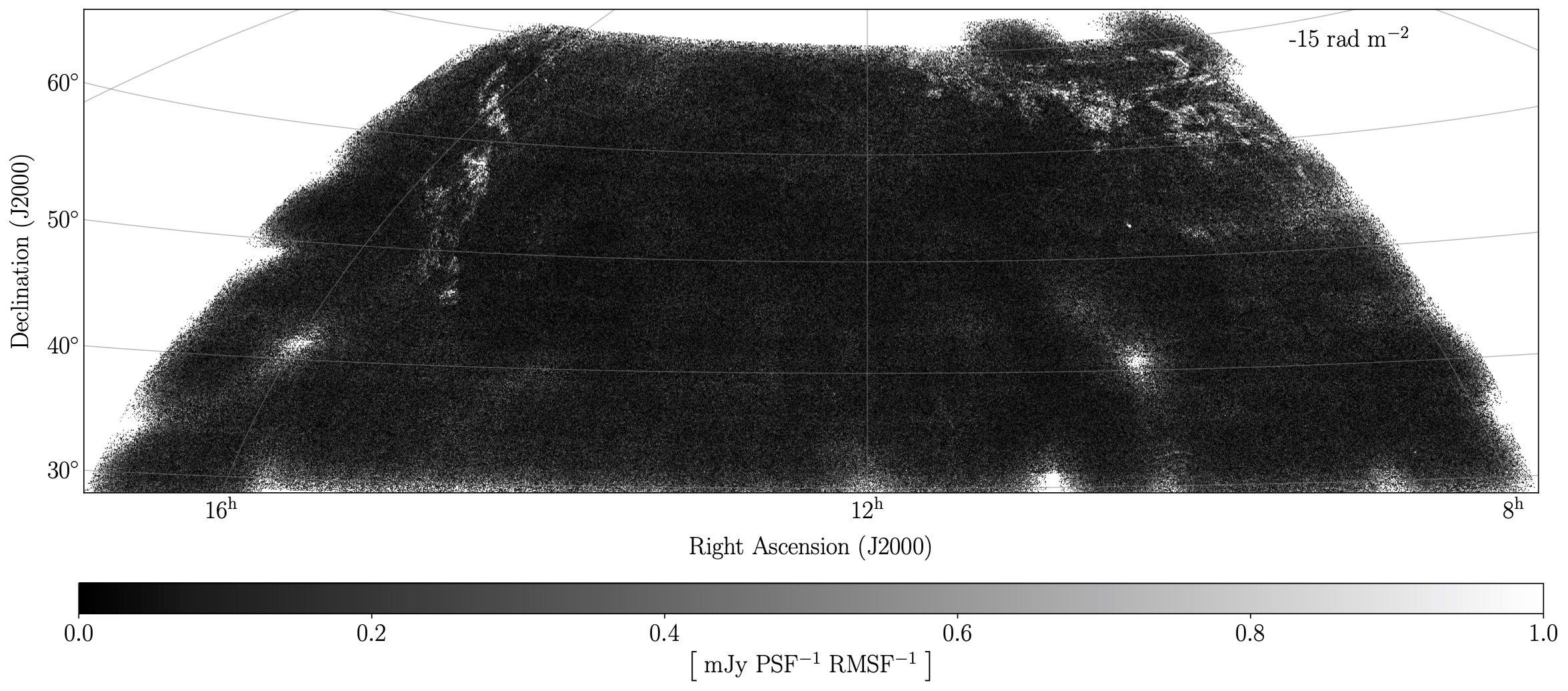}
      \caption{Slices from the mosaic Faraday cube at Faraday depths $-25, -20, -15~\mathrm{rad~m^{-2}}$. The images are dominated by Loop III emission.
                  }
         \label{-25_-15}
\end{figure}

\begin{figure*}[h]
   \centering
   \includegraphics[width=\textwidth]{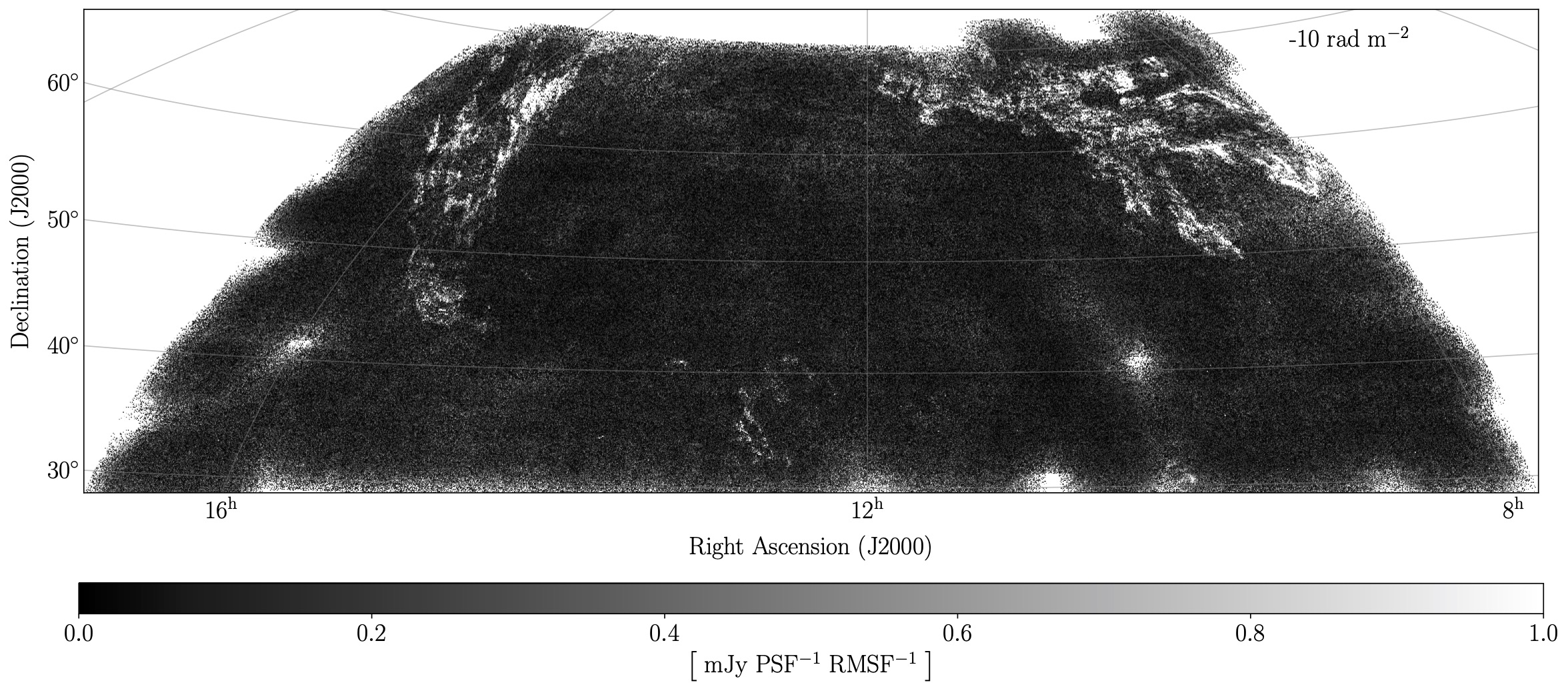}

   \includegraphics[width=\textwidth]{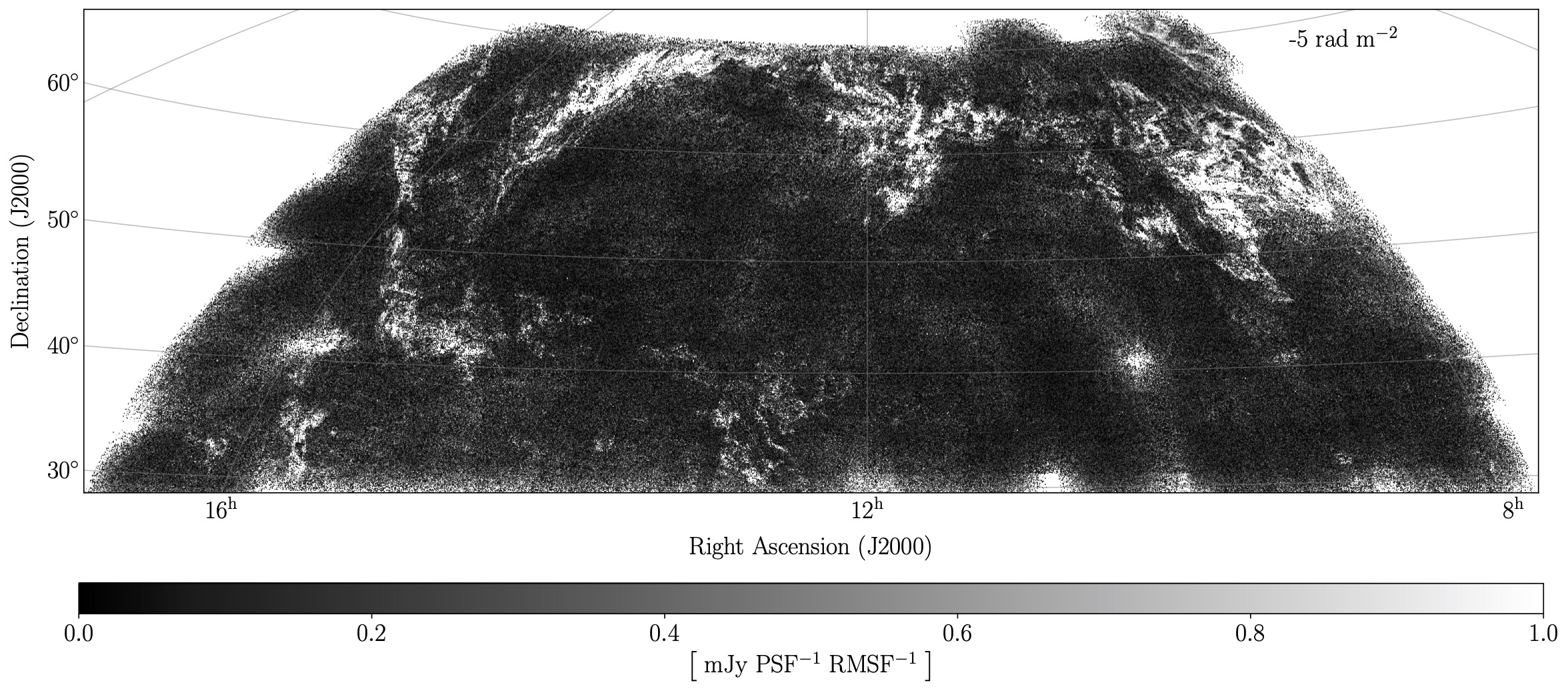}

   \includegraphics[width=\textwidth]{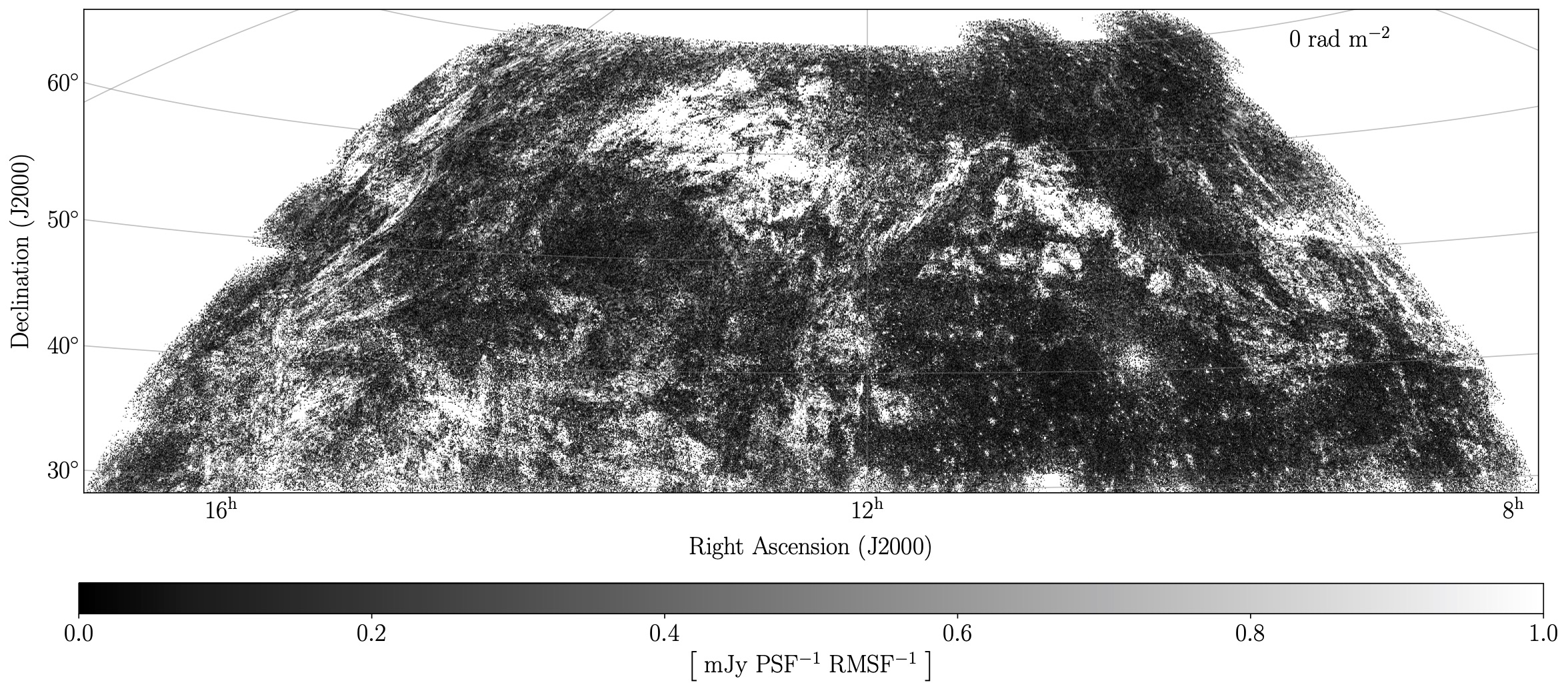}
      \caption{Slices from the mosaic Faraday cube at Faraday depths $-10, -5, 0~\mathrm{rad~m^{-2}}$. The images are dominated by Loop III emission.
                  }
         \label{-10_0}
\end{figure*}

\begin{figure*}[h]
   \centering
   \includegraphics[width=0.99\textwidth]{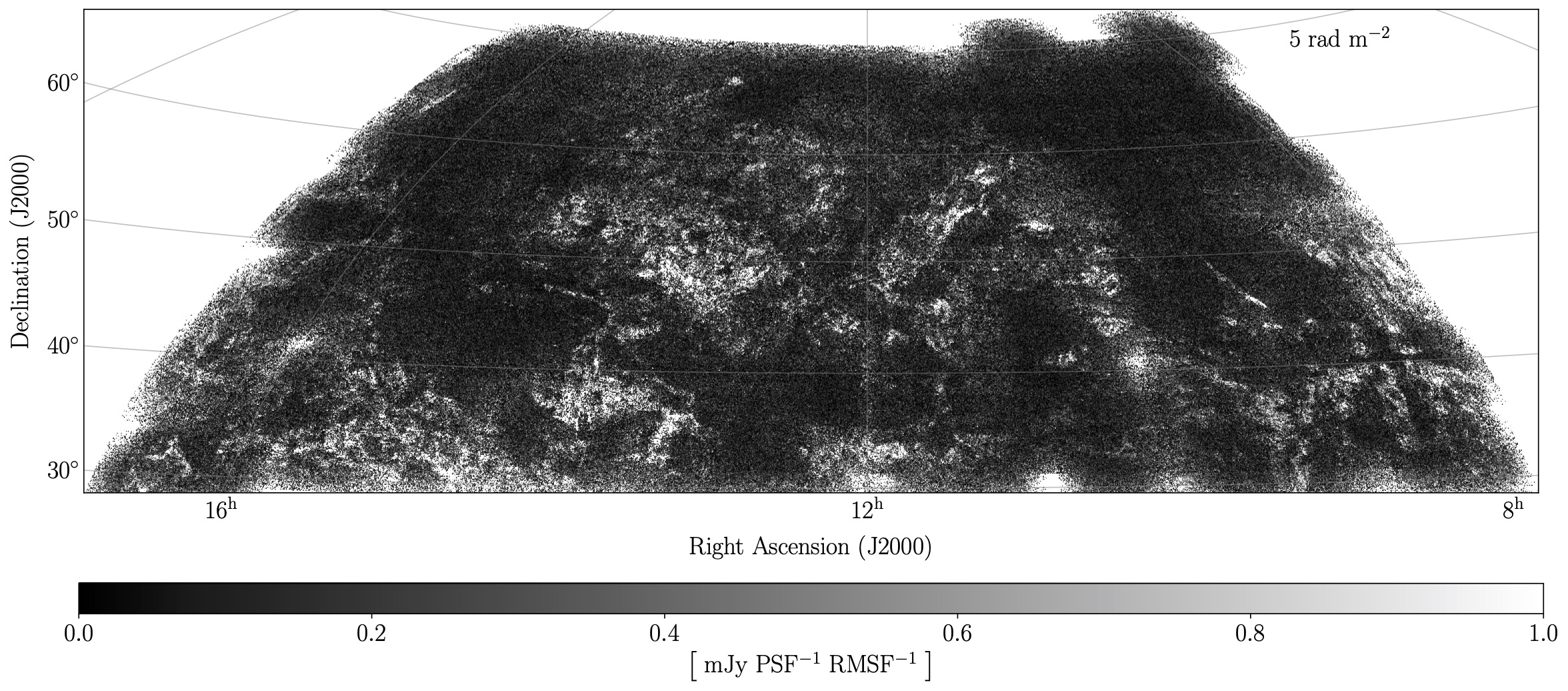}

   \includegraphics[width=0.99\textwidth]{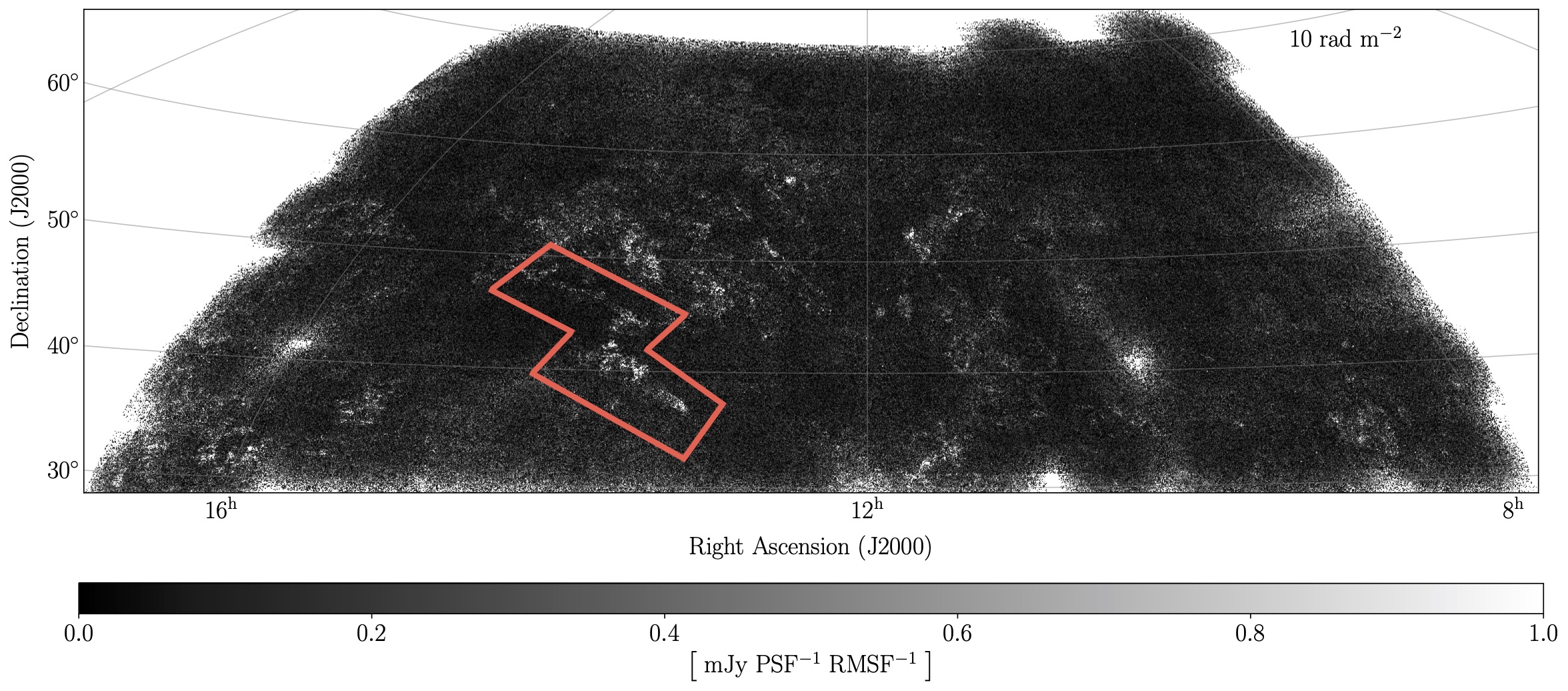}
         
   \includegraphics[width=0.99\textwidth]{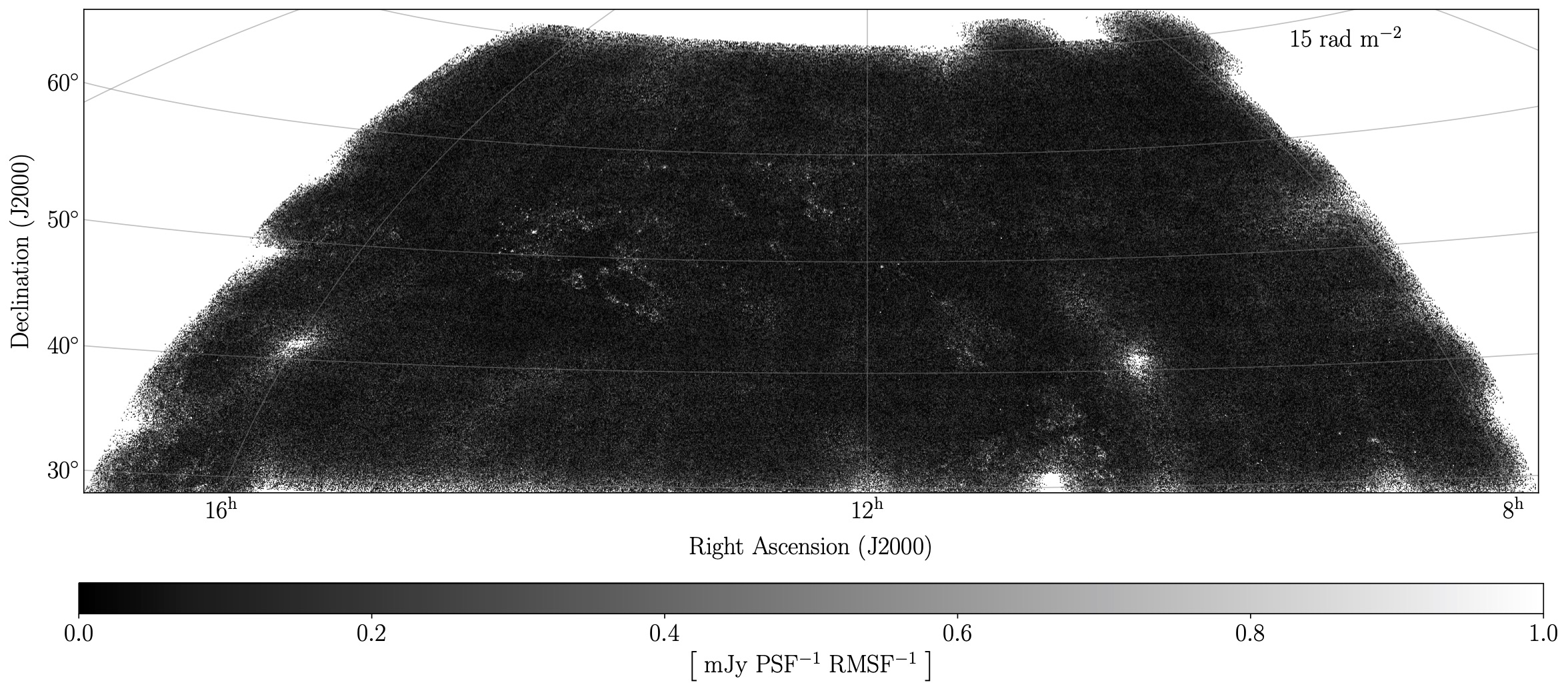}
      \caption{Slices from the mosaic Faraday cube at Faraday depths $5, 10, 15~\mathrm{rad~m^{-2}}$. Loop III emission in traces is seen at 5 $\mathrm{rad~m^{-2}}$. {While a part of the zig-zag structure can be seen in all of the panels, the area in which the zig-zag structure is best seen (at 10 $\mathrm{rad~m^{-2}}$) is roughly outlined in red.} At Faraday depths higher than 15 $\mathrm{rad~m^{-2}}$, we see no emission.
                  }
         \label{0_15}
\end{figure*}
}

\twocolumn{
\section{Toy models probing the multi-frequency comparison}\label{app:burn}

In this appendix, we describe the toy models used to interpret the correlations found in Sect.~\ref{sec:discussion} between data sets at different frequencies. In particular, we focus on the correlation between our LOFAR data and the total Galactic RM. 
For a given LOS, we consider three distinct cases, as shown by the sketches in Fig.~\ref{fig:toysketch}.

\begin{figure}[!h]
   \centering
   \includegraphics[width=\hsize]{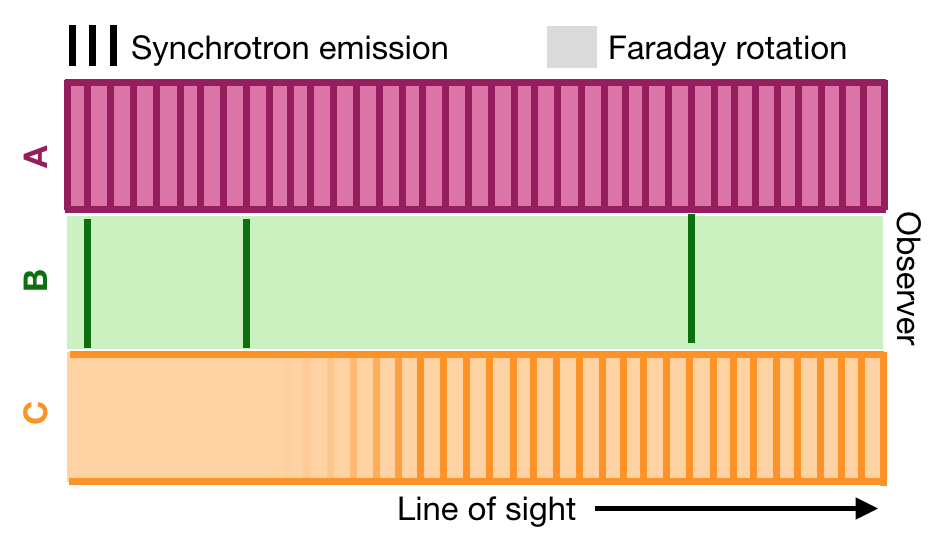}
      \caption{Sketches showing the three models (case A, B, and C) considered in our analysis to interpret the multi-frequency correlations discussed in Sect.~\ref{sec:discussion}.
              }
         \label{fig:toysketch}
\end{figure}

Case A is a Burn slab in which the full length of the LOS is affected by differential Faraday rotation as the result of a mixture of uniform synchrotron-emitting and Faraday-rotating media. {We assumed that we observe} along the mean magnetic field component, such that the equivalent value of the modelled Galactic RM (MGRM) is 20 rad m$^{-2}$. This value of MGRM is arbitrary, but large enough to certainly be a Faraday thick structure at the observed frequencies of LOFAR. The cosmic-ray electron density, the magnetic-field structure and strength, and the free-electron density are all homogeneous and uniform quantities along the LOS. 

Case B has the same MGRM as case A, but a different distribution of emission along the LOS. In this case {we considered} the synchrotron emission to be confined to one layer, or several distinct layers, so that emission and Faraday rotation are two separated processes. Each Faraday rotating medium between two different synchrotron emitting layers represents a Faraday screen. In this case we distinguish between three configurations. In the first one, B.0, one single emitting layer is affected by the full length of the Faraday rotating medium, or one single Faraday screen. In B.1, one or multiple discrete emitting layers are distributed across the full length of the Faraday rotating medium. In B.2, one or multiple discrete emitting layers are only distributed in the first half of the Faraday rotating medium towards the observer.  

Finally, case C is a modified version of case A in which the synchrotron emission is attenuated as a function of the distance from the observer. The MGRM in case C is the same as in the other two cases. 

In all three cases, we simulated LOFAR polarisation observations in the same range of frequencies as the one described in this paper and we performed Faraday tomography using {\tt{rm-synthesis}\footnote{\url{http://github.com/brentjens/rm-synthesis}}}\citep{brentjens05}. Thus, we produced Faraday spectra similar to those in Fig.~\ref{fig:spectratoy}, where cases A and C show two different configurations of Faraday thick structures of which we can only observe the edges \citep[e.g.][]{vaneck17}, and all the B cases show the position in Faraday space of each synchrotron emitting layer. From such spectra, we computed the first moments ($M_1$) and compared them to the value of MGRM. 

\begin{figure}[!h]
   \centering
   \includegraphics[width=\hsize]{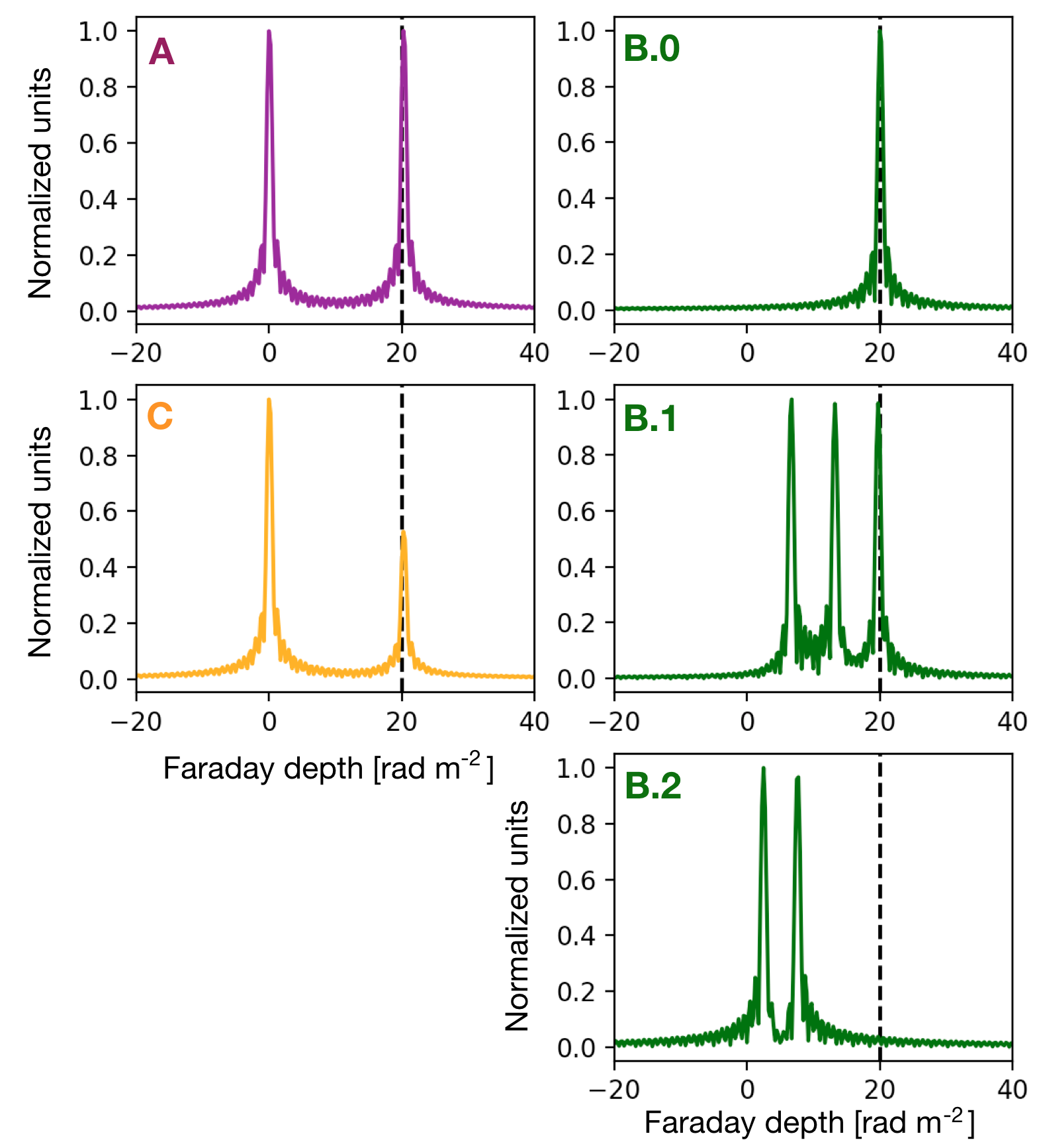}
      \caption{Modelled Faraday spectra of the three cases A, B, and C discussed in the main text. The vertical dashed line represents the value of MGRM = 20 rad m$^{-2}$.   
              } \label{fig:spectratoy}
\end{figure}

Figure~\ref{fig:toymodels} illustrates the ratios between MGRM and $M_1$ for each model. Only in case B.0 is MGRM/$M_1$ = 1, while MGRM/$M_1$ = 2 was obtained, as expected, in case A, where uniform differential Faraday rotation was modelled. The B.1 configuration {produced} values of MGRM/$M_1$ between 1 and 2, while both cases B.2 and C {returned} values of MGRM/$M_1$ > 2. Departures from the value of 2 are generally related to cases in which the Faraday rotating medium significantly extends over a longer portion of the LOS compared to the synchrotron emitting regions.    

\begin{figure}[!h]
   \centering
   \includegraphics[width=\hsize]{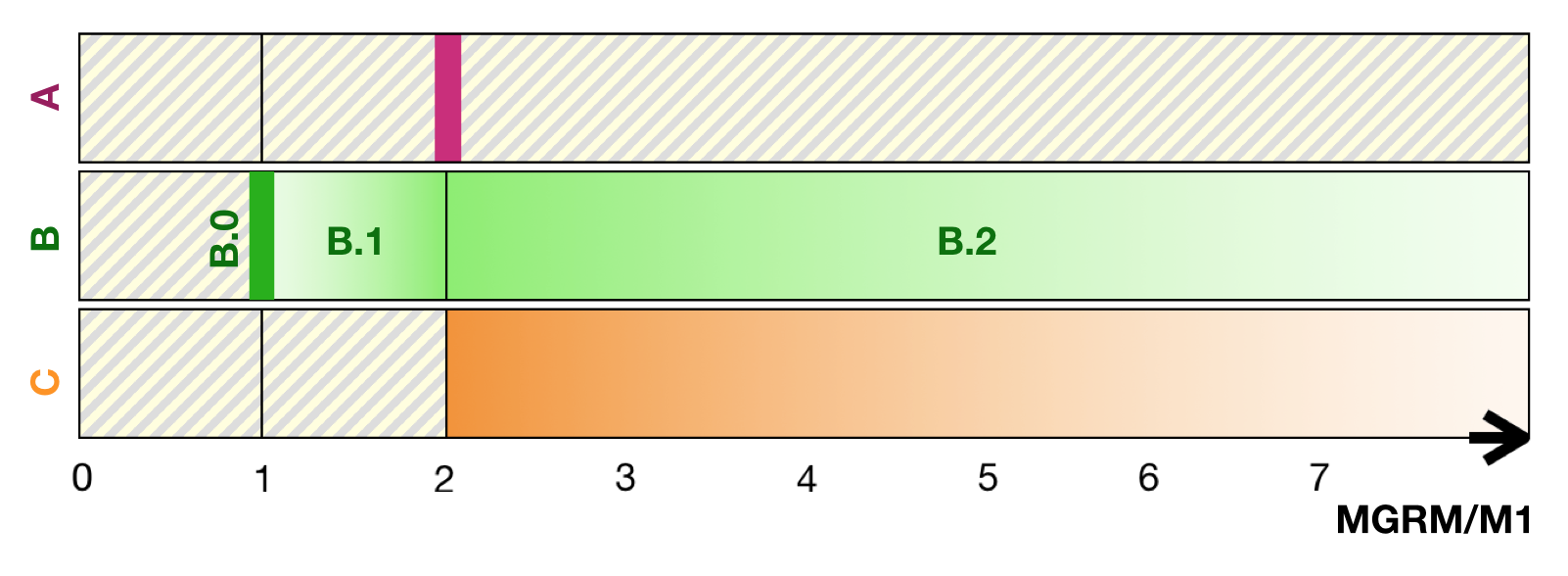}
      \caption{Diagram showing the ratio between the modelled Galactic RM and the value of the first moment for the simulated Faraday-tomographic data in the three cases (A, B, and C) described in the main text. Grey-shaded areas represent forbidden regions for each model.
              }
         \label{fig:toymodels}
\end{figure}
}
\end{appendix}
\end{document}